\documentclass[twocolumn,aps,longbibliography,superscriptaddress]{revtex4-1}

\usepackage[pdftex]{graphicx}  
\usepackage{amssymb}   
\usepackage{gensymb}
\usepackage{amsmath}
\usepackage{float}
\usepackage{xcolor}
\begin{document}

\title{Scaling in the recovery of urban transportation systems from special events}

\author{Aleix Bassolas}\email{aleix@ifisc.uib-csic.es}
\affiliation{Instituto de F\'isica Interdisciplinar y Sistemas Complejos IFISC (CSIC-UIB), Campus UIB, 07122 Palma de Mallorca, Spain}
\author{Riccardo Gallotti}\affiliation{Instituto de F\'isica Interdisciplinar y Sistemas Complejos IFISC (CSIC-UIB), Campus UIB, 07122 Palma de Mallorca, Spain}
\affiliation{Fondazione Bruno Kessler, Via Sommarive 18, 38123 Povo (TN), Italy}
\author{Fabio Lamanna}\affiliation{Instituto de F\'isica Interdisciplinar y Sistemas Complejos IFISC (CSIC-UIB), Campus UIB, 07122 Palma de Mallorca, Spain}
\author{Maxime Lenormand}
\affiliation{Irstea, UMR TETIS, 500 rue JF Breton, 34093 Montpellier, France}
\author{Jos\'e J. Ramasco}\email{jramasco@ifisc.uib-csic.es}\affiliation{Instituto de F\'isica Interdisciplinar y Sistemas Complejos IFISC (CSIC-UIB), Campus UIB, 07122 Palma de Mallorca, Spain}

\begin{abstract}
Public transportation is a fundamental infrastructure for the daily mobility in cities. Although its capacity is prepared for the usual demand, congestion may rise when huge crowds concentrate in special events such as massive demonstrations, concerts or sport events. In this work, we study the resilience and recovery of public transportation networks from massive gatherings by means of a stylized model mimicking the mobility of individuals through the multilayer transportation network. We focus on the delays produced by the congestion in the trips of both event participants and of other citizens doing their usual traveling in the background. Our model can be solved analytically for regular lattices showing that the average delay scales with the number of event participants with an exponent equal to the inverse of the lattice dimension. We then switch to real transportation networks of eight worldwide cities, and observe that there is a whole range of exponents depending on where the event is located. These exponents are distributed around $1/2$, which indicates that most of the local structure of the network is two dimensional. Yet, some of the exponents are below (above) that value, implying a local dimension higher (lower) than $2$ as a consequence of the multimodality and multifractality of transportation networks. In fact, these exponents can be also obtained from the scaling of the capacity with the distance from the event. Overall, our methodology allows to dynamically probe the local dimensionality of a transportation network and identify the most vulnerable spots in cities for the celebration of massive events.
\end{abstract}
\maketitle

\section*{Introduction}

The recent boost in accessibility to data on human mobility \cite{Gonzalez2008} and infrastructures facilitates, more than ever, the study of human movement and transportation systems \cite{Barbosa2018}. Human mobility covers a wide range of scales as a reflection of several transportation modes, yet the increase of urban population makes the management of inter-urban infrastructures a critical subject. For many years now, the dichotomy between private on-road and public transportation has dominated urban mobility. However, in a society increasingly concerned about climate change, and with road traffic accounting for more one fifth of the total emissions in the US \cite{US}, fostering the use of public transportation is nowadays a priority. Among the factors driving the public transportation use \cite{Paulley2006,Balcombe2004}, improving travel times and keeping congestion under control are fundamental. 


In the recent years, complex network theory has been proved as an useful tool to model and understand dynamics of real world systems like epidemic spreading \cite{vespi}, transportation systems \cite{Fleurquin} or social interactions \cite{twitter}. Transportation systems (e.g., road, rail, bus, air) can be naturally mapped into networks, which helps to unveil systemic features \cite{boccaletti,newman}. Similarly to power grids, transportation networks are intrinsically spatial \cite{Barthelemy2011,Louf2013}, space is inseparable from their topological and dynamical properties. The first works on public transportation focused mainly on their topology \cite{Latora2002,Sen2003}, irrespective of the travel times between nodes and the spatial component. The next step was to introduce weights as a representation of travel times or distances between nodes \cite{barrat,ramasco}. Those weights become even more relevant when more than one mode--with different characteristics--are intertwined, as in the case of public transportation \cite{navigability,user1}. The recent introduction of the multilayer framework \cite{kurant,kivela} and its development \cite{reductibility} offers new tools to deal with a more complex reality, providing a unique framework to model multimodal transportation. From a theoretical perspective, it was shown that multilayer networks can, by themselves, give rise to congestion \cite{Sole2016} and that an increase of link speeds or capacities does not necessarily improve network performance \cite{Manfredi2018}. The onset of congestion in multiplexes (multilayer networks in which each node represents the same entity in different layers) have been studied as a function of the number of layers and degenerate paths \cite{Sole2019}.
From a practical perspective, several works studying multimodal transportation networks as multiplexes have been published:  From works where every transportation mode is a single layer \cite{GreatB,gallotti2014anatomy,navigability} to others using a layer for each line \cite{Aleta2017,user1}. 

Understanding the emergence of congestion and delays requires, however, a step beyond the static structure of the network. First of all, in both private and public transportation the capacities of lines, roads and vehicles are of wide variety, and need to be taken into account to evaluate congestion and vulnerability \cite{ccolak2016understanding,Chodrow2016,Manfredi2018}. Secondly, the transportation demand in cities is by no means homogeneous. Neither the origins and destinations of trips are equiprobable nor the number of travelers is constant in time. In fact, the interest of planners in the spatio-temporal characteristics of demand is almost as old as transportation research \cite{Mcfadden1974}. Luckily, nowadays we have new sources of data to understand urban mobility as a cheaper alternative to the usual surveys \cite{Lenormand2014,Colak2015,Jiang2016}. While most works focus on shortest paths to infer the routing of individuals, the way citizens choose their route is far from trivial \cite{De2015,Lima2016}, specially in the congested phase in which their routes might be readapted \cite{Sole2016b}.

Whereas transportation infrastructures are prepared for the daily demand, they might fail to manage shocks and disruptions \cite{navigability,Silva2015}. Particularly, concerts, sport events or massive demonstrations are recurrent in metropolises, gathering huge crowds in localized places. Such a peak of demand in a small area affects the whole public transportation system, yet this situation has attracted limited interest so far \cite{Xu2017}. In this work, we examine how the delay of individuals leaving an event depends on where it takes place and the number of attendees. Such dependence turns out to follow power-law-like (scaling) relations. The average delay in public transportation networks of eight worldwide cities (Amsterdam, Berlin, Boston, Madrid, Milan, New York City, Paris and San Francisco) has been numerically analyzed, showing that the scaling holds and the exponents can be explained by a newly proposed local dimension metric taking the capacity of lines into account. Thereafter, we prove the relation between the scaling and the network dimension by solving the model analytically in regular lattices.

\section*{Methods} 

\subsection*{Model description} 

Inspired by previous works on information transmission in computer systems \cite{Echenique2005}, we build a model to simulate how individuals move through a public transportation system using vehicles with limited capacity. The main ingredients are {\it i)} the multilayer network, {\it ii)} vehicle mobility, {\it iii)} individuals trip planning and execution, and {\it iv)} the distribution of origins and destinations of individuals (transportation demand). 

\begin{figure}
\centering
\includegraphics[width=\columnwidth]{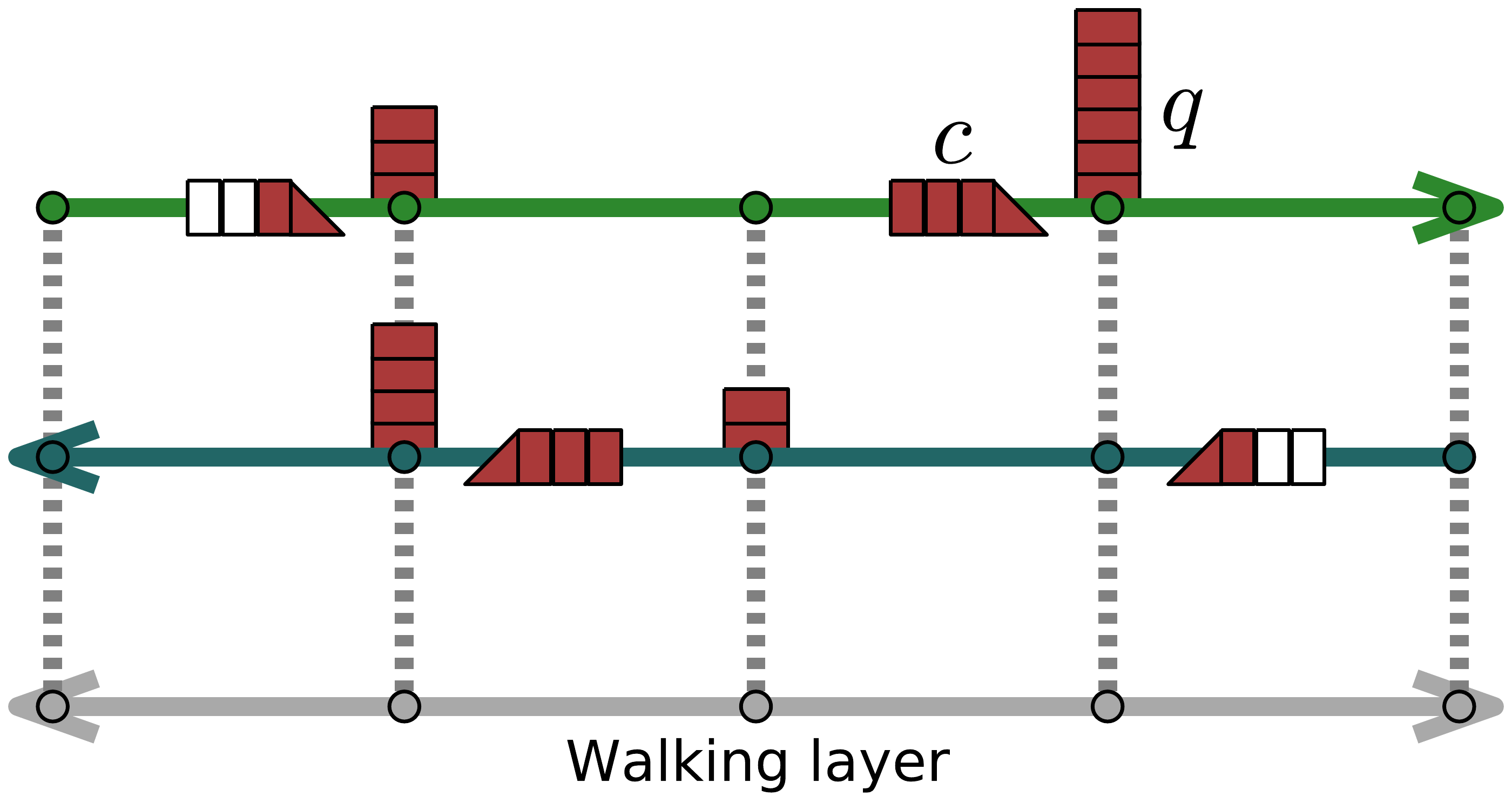}
\caption{Sketch of the multilayer structure of our model. The two top layers are the transportation lines through which vehicles move. Individuals wait at each node generating a queue $q$ until a vehicle arrives and they are able to board if the vehicle is below full capacity $c$. Below the transportation lines, in gray, a the walking layer with infinite capacity connected to the transportation lines layer through inter-layer links.}
\label{Figure1}
\end{figure}

The public transportation system of cities is, in general, formed by different modes, each one with its own characteristics in terms of speed, period,
and capacity. The line period (or headway) is defined as the lapse of time between consecutive departures of vehicles in the same route. Similarly to \cite{Aleta2017,user1}, we model the public transportation system as a multilayer network where every direction of each line is represented by one layer. The urban space is divided into a grid of cells, with 
size $400m \times 400m$. The stops of every line are assigned to the cell where they are located, so that the nodes of a layer correspond to cells hosting stops of the considered directed line. If more than one stop falls in the same cell, they are merged into a single node. Directional weighted intra-layer links are then created between nodes hosting consecutive stops. The weight corresponds to the travel time, calculated as the distance between connected cells centroids divided by the average speed of the transportation mode. Within this framework, all nodes have one incoming and one outgoing link except those containing the beginning and end of the line. In the multilayer network, we call a \textit{location} the set of nodes associated to the same cell, and 
all nodes within the same location 
are
connected through inter-layer links. Individuals can change line along these links with an average (walking) time penalty of $30$ seconds. Additionally, besides all lines, we introduce a further walking layer with infinite capacity so that every cell is also a node in this layer. The nodes of the walking layer are connected within a certain radius ensuring its full connectivity, and travel times are calculated as the Haversine distance divided by a walking speed of $5$ km/h. A sketch of how both directions of a line would be connected within our framework is depicted in Fig. \ref{Figure1}. For simplicity and visualization purposes, the figure depicts a 1D lattice.  

The vehicles leave the line origin location with a fixed period and they go through all route nodes until the terminal.
Since line periods vary throughout the day, we select the best case by adopting the schedule of $8$am (rush hour). In a more realistic scenario, transportation managers may reinforce the lines as a response to an event celebration. We do not have this detailed information so we neglect this effect.

The actions of individuals have two phases: one of planning and one of execution. As the model is adaptive, these phases can be repeated until she/he arrives at the final destination. During the trip execution, individuals can move freely through the walking layer but need to use a vehicle to travel along links in the transportation lines. 
In case of line change, the waiting time is determined by the vehicle movements. At every node, we consider a queue $q$ of individuals waiting for a vehicle. They board in order (regulated on a "{\it first come, first served}" (FIFO) basis) until filling the vehicle or emptying the queue. Each vehicle has a capacity $c$ given by the transportation mode and it is considered equal for all the vehicles in the same line. 

\begin{figure}
\begin{center}
\includegraphics[width=\columnwidth]{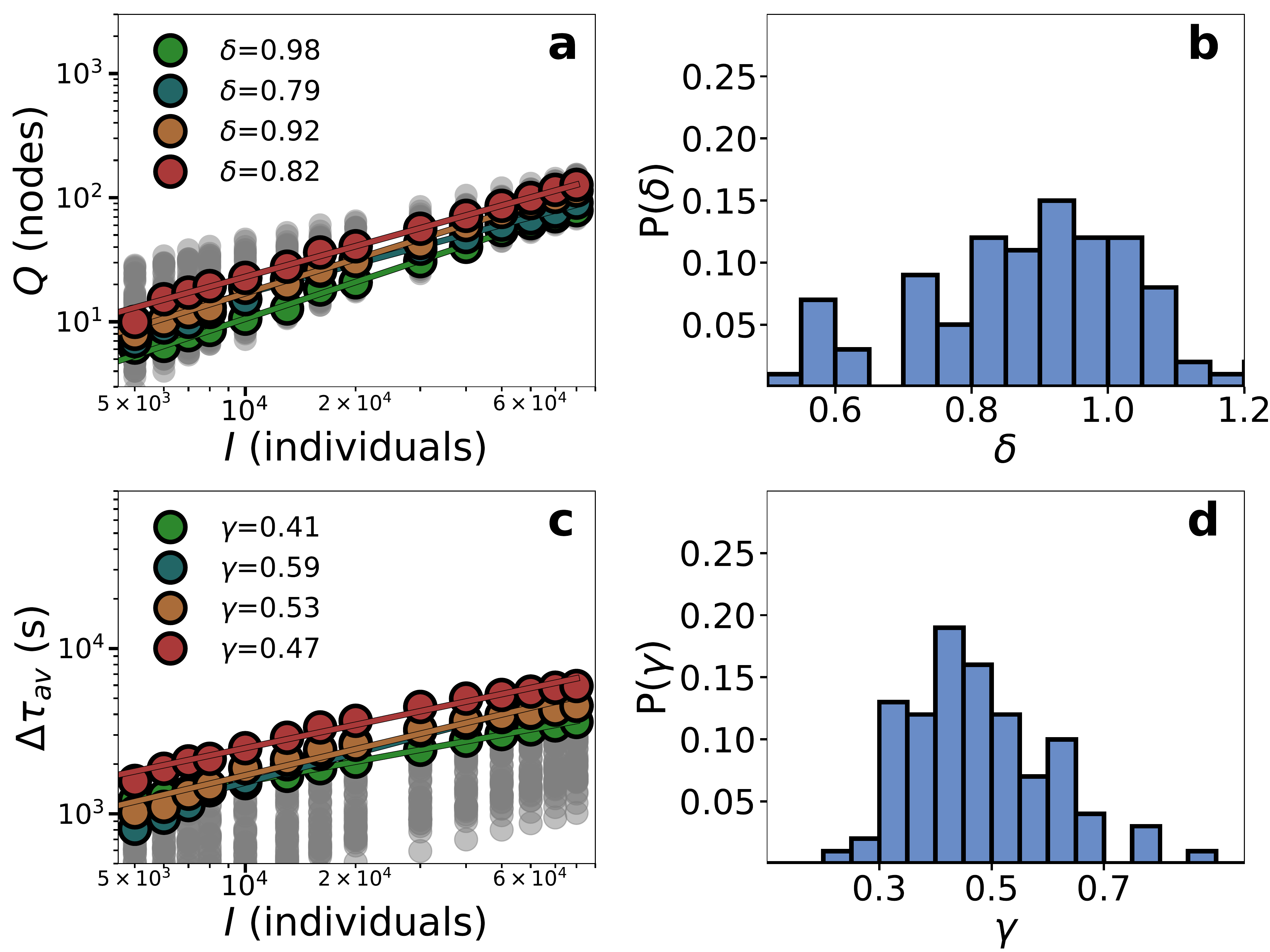}
\caption{Scaling for the event individuals in Paris. \textbf{a} Congested nodes as a function of $I$. \textbf{b} Distribution of the exponents of the scaling of congested nodes. \textbf{c} Average delay vs $I$. \textbf{d} Distribution of the exponents of the scaling of the average delay.
\label{Figure2}}
\end{center}
\end{figure}

The routing protocol used by individuals for planning is adaptive with local information. In the absence of congestion, individuals follow the temporal optimal path of the static multilayer network calculated by the Dijkstra algorithm \cite{Ahuja1988}. If there are line changes, they estimate an additional waiting time of half the new line period. 
Their route is only recalculated when a congested node, whose queue is larger than the vehicle's capacity, is reached. In the case of rerouting, the link travel-time for an individual $i$ attempting to board is updated with the expected waiting time given by 
\begin{equation} \label{eq1}
t_{wait,i}=\left(\frac{1}{2}+\left[\frac{q_i}{c}\right]\right) \, f, 
\end{equation}
where $q_i$ is the queue observed by individual $i$, $c$ is the capacity of the line and $f$ is the period, 
i.e. the time between consecutive vehicles. The square brackets $[q_i/c]$ represent the integer part of the ratio. Simply put, the new inter-layer weight will be the period multiplied by the number of vehicles necessary to empty the queue plus the normal waiting time of $f/2$.

The transportation demand is estimated using Twitter data to construct origin-destination (OD) matrices if the model applies to cities. For each individual, the residence and work places are assigned to the most common locations between $8$pm and $8$am and between $8$am and $8$pm, respectively. Users with less than $10$ tweets and traveling faster than $300 \, km/h$ are disregarded. The flows between origins and destinations of trips are then aggregated from the distribution of work and residence places of all the valid users. In the case of lattices, the trip demand is uniform. 

Throughout the work, individuals are separated in two: those participating at the event and those in the background. The trip demand of the background individuals is generated at a rate $\rho$ with an origin and a destination given by the OD matrix extracted from Twitter. We ensure that the injection rate is low enough so that delays are not generated in the standard operation scenario. A number of individuals $I$ is introduced at the place and time in which the massive event occurs, with their destinations chosen according to the residence distribution in the OD matrix. Simulations run until all the $I$ individuals reach their destinations, and the average delay per individual $\Delta\tau_{av}$ is then calculated. Given the real travel time of the individual $\tau_{real}$ and the expected $\tau_{op}$, measured with the optimal path, the average delay taken over all the delayed individuals is calculated as $\Delta\tau_{av} = \langle \tau_{real}-\tau_{op} \rangle$.

\section*{Results} 

\subsection*{Scaling in cities}

\subsubsection*{Event individuals}

\begin{figure}
\begin{center}
\includegraphics[width=\columnwidth]{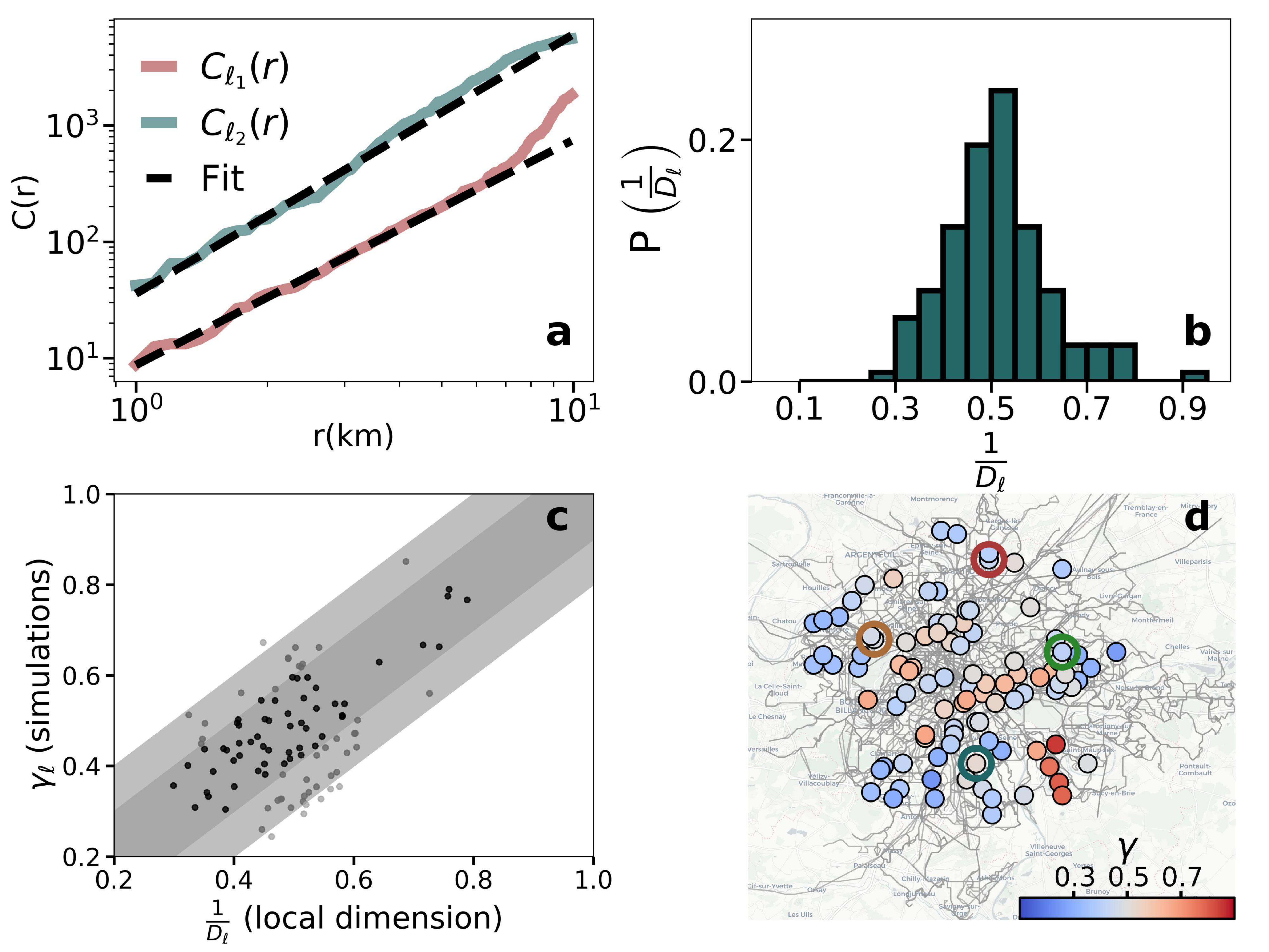}
\caption{Calculation of the local dimension $D_\ell$ in the multilayer network of Paris and comparison between $D_\ell$ and $\Delta\tau_{av}$. \textbf{a} In solid line the scaling of the capacity $C(r)$ as a function of the distance to the event in two locations and the power law fit in dashed line. \textbf{d} Distribution of the inverse of the local dimension $D_\ell$ in the same locations where simulations were performed (see Fig. \ref{Figure4}). \textbf{c} Inverse of the local dimension $D_\ell$ vs the exponent obtained from the simulation  $\gamma_\ell$ in the same locations, the dark and light gray bands correspond to margins of error of $0.1$ and $0.2$, respectively. \textbf{d} Map of the scaling exponents of the average delay. The empty circles in the maps mark the locations of the scaling shown in Fig. \ref{Figure2}.\label{Figure3}}
\end{center}
\end{figure}

We investigate first how the average delay per individual $\Delta\tau_{av}$ and the number of congested stops $Q_T$ scale with the number of individuals in the event $I$ for eight empirical public transportation networks. Starting with $Q_T$, Fig. \ref{Figure2}a displays its scaling $Q_T \sim I^{\delta}$ in four Paris locations. Even though there is  only a little more than one decade, the curves in log-log are straight and the scaling exponent is similar across locations. Analyzing the distribution of exponents in more detail (Fig. \ref{Figure2}b), we observe that the distribution of $\delta$ is centered around {\bf $0.95$} and it could be compatible with a linear behavior. In contrast to the scaling, the number of congested nodes does depend on the event location.

Looking at the delay of the individuals attending the event, Fig. \ref{Figure2}c displays the relation between $\Delta\tau_{av}$ and $I$ in the same four locations. As before, the values of the delays depend on the location (the curves show the same order as in Fig. \ref{Figure2}a, more congestion implies more delay). The curves can be approximated with scaling relations $\Delta\tau_{av}\sim I^{\gamma_{\ell}}$, yet the exponents $\gamma_\ell$ in every location $\ell$ show larger relative variability. The relationship is sub-linear in all cases with most of the exponents between $0.3$ and $0.7$ and centered around $0.5$ (Fig. \ref{Figure2}d). As we will show later, the exponents are related to the local characteristics of the network and the inverse of $\gamma$ can be seen as a local dimension centered around $2$ as expected from a spatial network. Dynamical features of complex networks have been previously related to their dimension \cite{Daqing2011,Gallotti2016}. The role of the local dimension can be further explored by introducing an alternative metric inspired by \cite{Daqing2011} but taking the network capacity into account. If the total capacity $C(r)$ in persons per second inside a circle of radius $r$ is calculated around the event location $\ell$, we find that $C_{\ell}(r)$ grows as  $C_{\ell}(r)\sim r^{D_{\ell}}$. The exponent $D_\ell$ is connected to the number and line types crossing close to $\ell$ and in lattices it corresponds to the network dimension. It is important to stress that $D_{\ell}$ is measured in the physical space, not in the network, and we take into account the capacity instead of the number of nodes usually used while measuring network dimension. An example of how $D_\ell$ is estimated from the empirical network of Paris is shown in Fig. \ref{Figure3}a. For consistency, we should expect that $D_\ell \approx 1/\gamma_{\ell}$ in every  $\ell$. In fact, the distribution of exponents measured in these two ways are very similar (see Fig. \ref{Figure3}b and Fig. \ref{Figure2}d). As can be seen in Fig. \ref{Figure3}c, $55\%$ of the points fall within the dark gray band and $93\%$ within the dark and light bands, which correspond to an error of the exponent of $0.1$ and $0.2$, respectively. The spatial variability can be seen in Fig. \ref{Figure3}d, where the exponent $\gamma$  obtained for a set of $100$ event locations homogeneously distributed across Paris is displayed. For public transportation systems, the local environment of each location is mostly unique leading to a highly heterogeneous map.

\begin{figure}
\begin{center}
\includegraphics[width=\columnwidth]{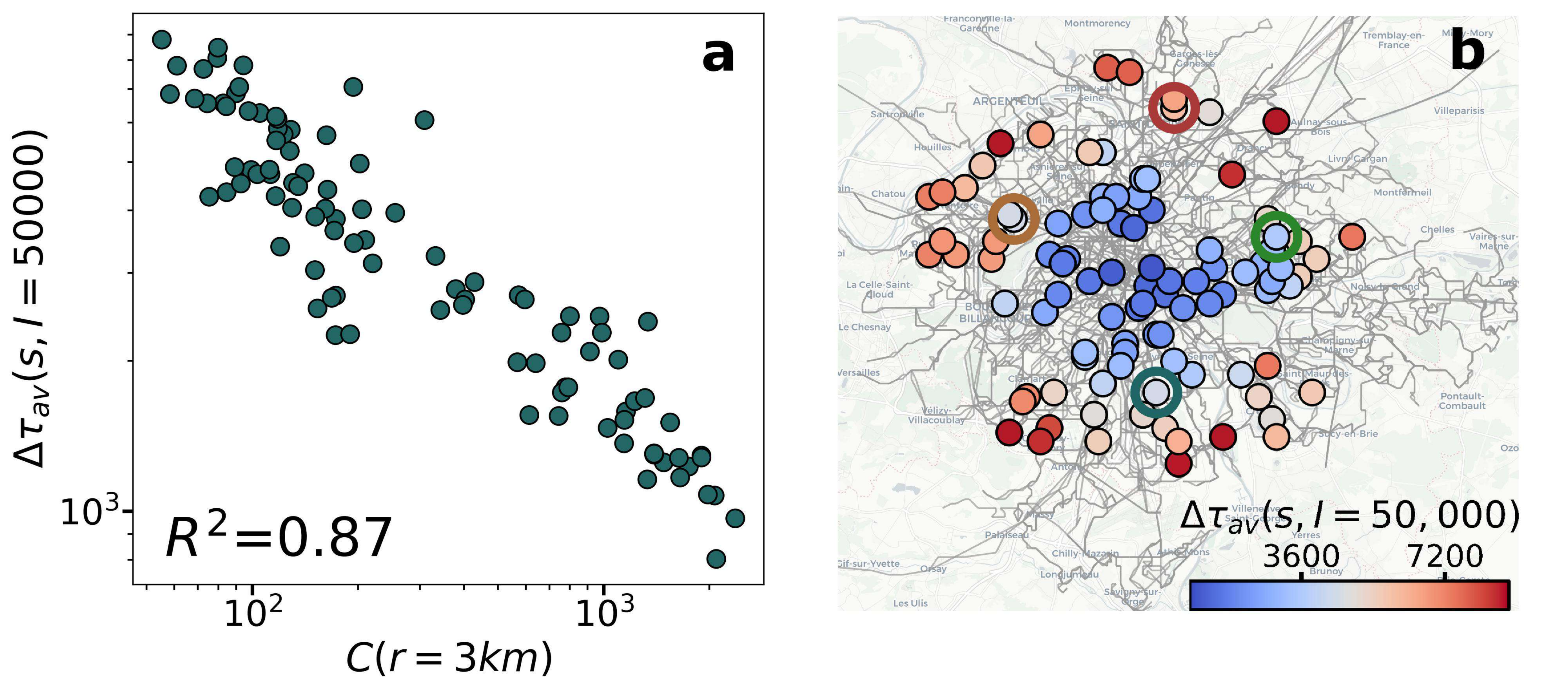}
\caption{Spatial distribution of delays. \textbf{a} $\Delta\tau_{av}$ for an event of $50,000$ individuals as a function of the total capacity within a radius of $3$ km. \textbf{b} Map of the average delay for an event of $50,000$ individuals. The empty circles in the maps mark the locations of the scaling shown in Fig. \ref{Figure2}.
\label{Figure4}}
\end{center}
\end{figure}

While the exponent $\gamma$ controls the growth of $\Delta\tau_{av}$ for $I \rightarrow \infty$, in real events the value of $I$ can be large but it is finite. We inspect thus the $\Delta\tau_{av}$ for an event gathering $50,000$ individuals in the previous $100$ locations. As shown in Fig. \ref{Figure4}a, the total capacity within a radius of $3$ km from the event location turns out to be a reasonable proxy for the value of $\Delta\tau_{av}$. The same relation can be observed for the other seven cities (see Appendix Fig. S10). Note that the spatial distribution of $\Delta\tau_{av}$ does not resemble the one of the exponents $\gamma_\ell$ (Fig. \ref{Figure4}b). This is due to the fact that when $I$ is fixed, the total capacity at the event location plays a dominant role. Nevertheless, if $I$ increases, the growth of the capacity around the event location is the most relevant variable. In this context,  locations close to the city center have access to a higher diversity of lines, while in the periphery the service is more limited (Fig. S10 in the SI). Consequently, the delays are lower in the central areas and higher in the periphery.
 
Similar results to those shown here for Paris are found in all seven other cities considered in the study (details can be found in the SI Section S4, Figs. S6-S9).

\begin{figure*}
\begin{center}
\includegraphics[width=\textwidth]{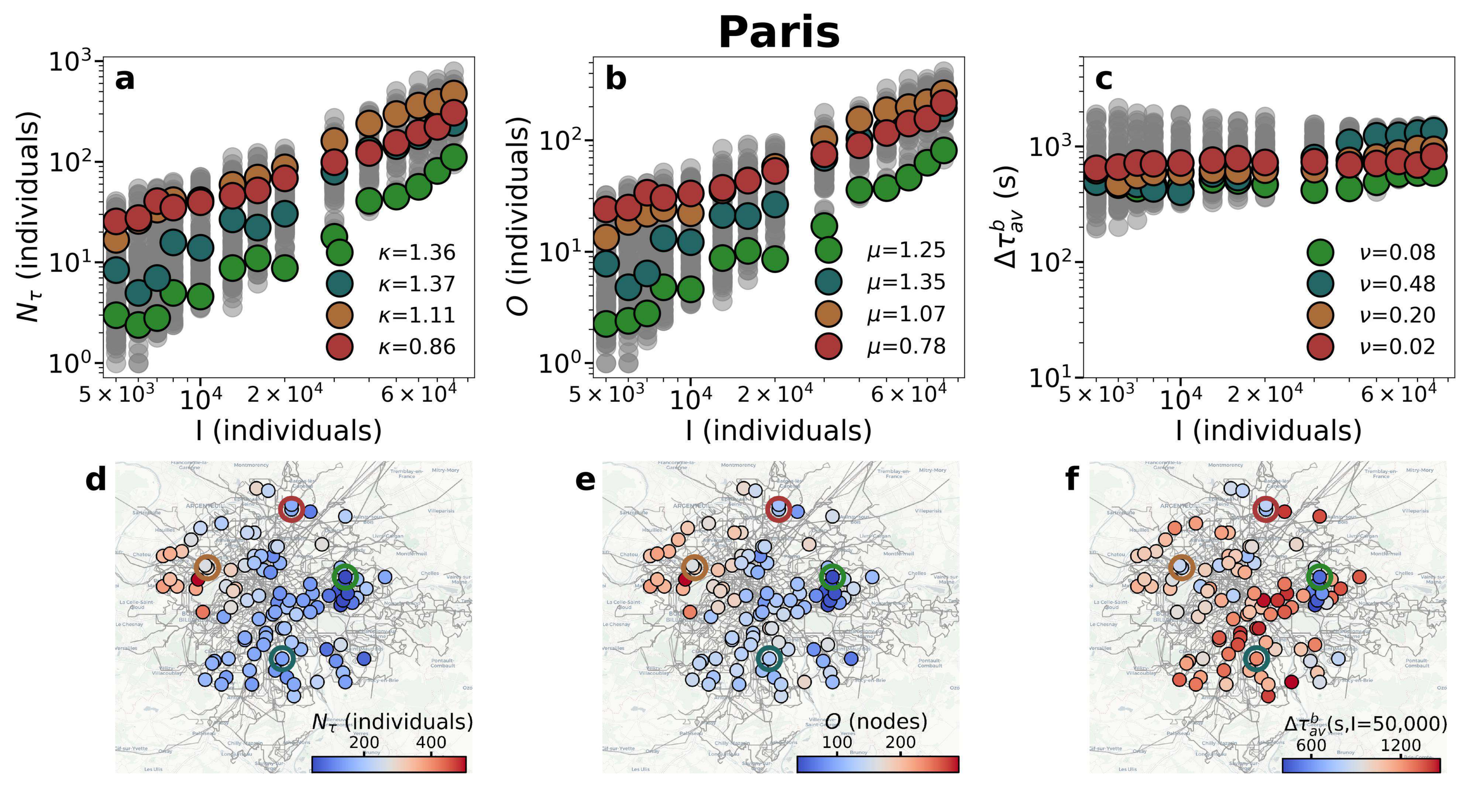}
\caption[Scaling for the background individuals in Paris]{Scaling for the background individuals in Paris. Scaling of the \textbf{a} delayed individuals, \textbf{b} origins affected and \textbf{c} average delay with the number event individuals. Map of the \textbf{d} delayed individuals, \textbf{e} origins affected and \textbf{f} average delay for an event of $50,000$ individuals. The empty circles in the maps mark the locations of the scaling shown in \textbf{a}, \textbf{b} and \textbf{c}. The empty circles in the maps mark the locations of the scaling.
\label{FigureS6}}
\end{center}
\end{figure*}

\subsubsection*{Background individuals}

Beyond the individuals participating in the event, the rest of citizens may also suffer delays as a consequence of the generated congestion. Scaling relations appear as well with $I$ in the case of the average delay per background individual, the number of delayed individuals and the number of affected trip origins. As before, the exponents are related to the local network dimension. To measure the effect of an event in the background individuals, we propose three metrics: the number of delayed individuals $N_\tau$, the number of unique origins of the individuals delayed $O$ and their average delay $\Delta \tau^b_{av}$. In Fig. \ref{FigureS6}, we display the scaling of those three metrics for Paris. In Fig. \ref{FigureS6}a, we show the number of delayed individuals as a function of $I$ with a super-linear scaling relation with an exponent around $1.5$. In Fig. \ref{FigureS6}b, we show the scaling of the number of origins affected, whose exponent falls between $1$ and $1.5$, yet it depends on the injection rate $\rho$. If $\rho$ is low, the number of congested nodes is low as well. Since the origin of background trips is random, in this regime it may happen that no trip is affected and the number of origins is lower than that of congested nodes. While if $\rho$ is high, all congested nodes are affected origins (which scales linearly with $I$). We observe in Fig. \ref{FigureS6}c, that the scaling of the average delay induced is sub-linear, similarly to the delay of the event individuals. This quantity is more noisy than the two previous metrics due to the small number of delayed individuals and their different origins.

Finally, we examine the effect of the event location on the background with a fixed $I$. In Fig. \ref{FigureS6}, the number of individuals delayed, their average delay and the number of affected origins are shown. As expected, events located in the periphery impact more individuals and origins, and produce higher delays. Even though the effect is similar as for event participants (Fig. \ref{FigureS6}f), the locations more affected are different. Event located in the west side of the city seem to affect more origins and individuals than other locations. Similarly, Figures S11-S17 of the SI display the results of the scaling for the individuals in the background.

\subsection*{Scaling in lattices}

In empirical networks, our analysis is constrained to a numerical perspective. We can, however, analytically prove in lattices that the scaling exits and show how the exponents are related to the network dimension,  

\subsubsection*{Event Individuals}
\paragraph*{1D lattices.}

\begin{figure*}
\centering
\includegraphics[width=\textwidth]{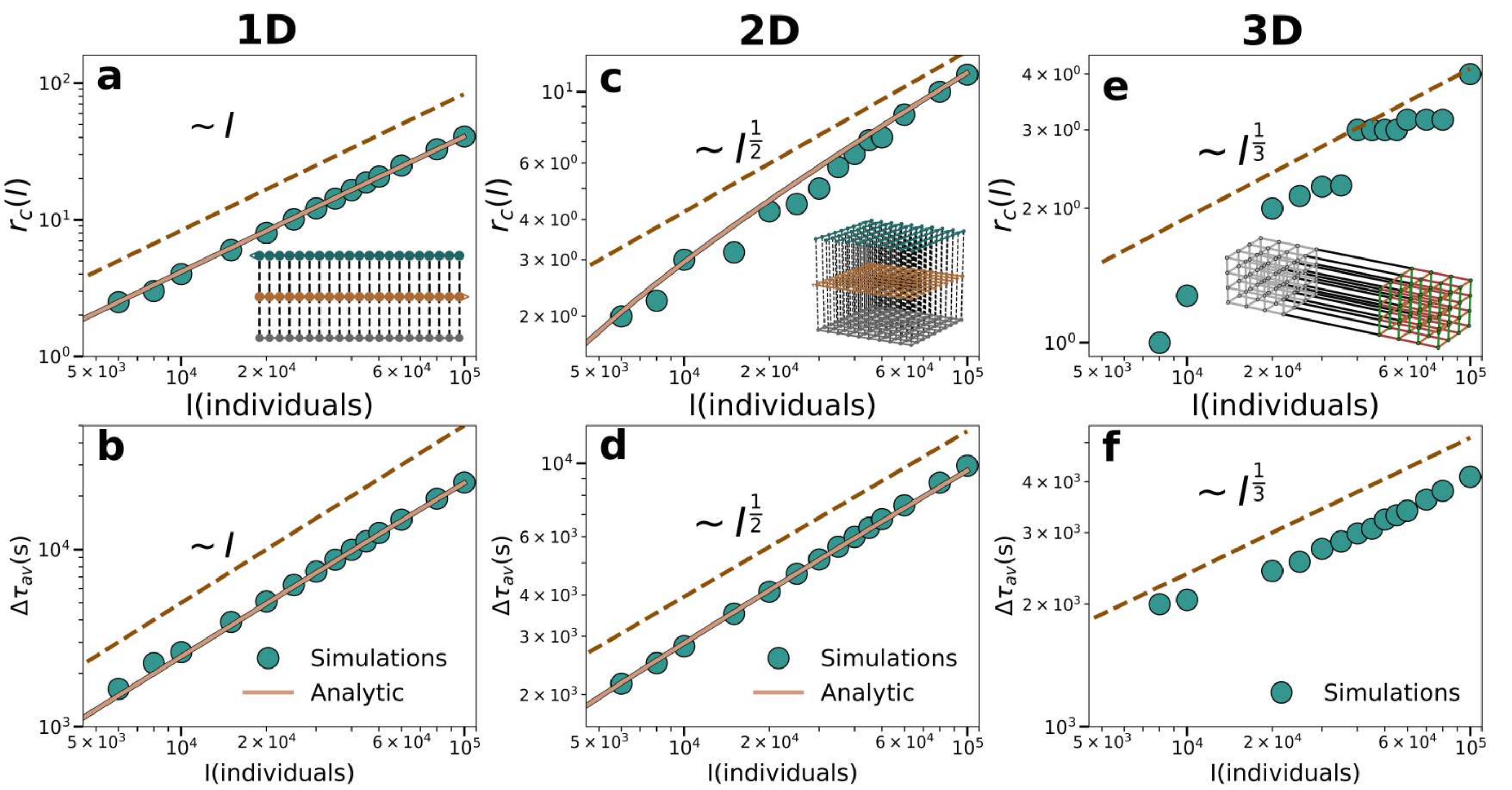}
\caption{Results in lattices with the event located at the center. In 1D, simulations were performed in a lattice of size $N = L=1500$ nodes, capacity $c=600$ persons and period $f=600$ seconds; In 2D, with $N = L^2 = 70^{2}$ nodes, $c=80$ persons and $f=600$ seconds in all the lines; And, in 3D with $N = L^3 =25^{3}$ nodes, $c=600$ persons and $f=600$ seconds. \textbf{a}, \textbf{c} and \textbf{e} radius of congestion $r_c$ depending on the attendees of the event $I$. \textbf{b}, \textbf{d} and \textbf{f} scaling of the average delay of agents depending on $I$. The dashed lines serve as a guide to the eye and the solid orange lines are the full analytical solutions provided in Sections S1 (1D) and S2 (2D) in the SI. All results correspond to the situation without background mobility and the inset shows a sketch of the multilayer network. }
\label{Figure5}
\end{figure*}

We start by building a network composed of only two lines going in opposite directions and a walking layer with infinite capacity connecting adjacent nodes (inset of Fig. \ref{Figure5}a and Fig. \ref{Figure1}). In such situation, individuals have only two alternatives: waiting for the next vehicle or walking towards the next node. The optimal choice depends only on the queue at the current node and the parameters of the system. Taking into account the travel time to the next node using a vehicle $t_1$ and by walking $t_2$, we can calculate from \eqref{eq1} the critical value of $q_i$ ($q^{\star}$) above which walking becomes the best option as 
\begin{equation}\label{qstar}
q^{\star}=  \left[\frac{t_2-t_1}{f}+\frac{1}{2}\right] \, c .
\end{equation}
Recall that $[.]$ stands for integer part and note that this equation is independent of the lattice dimension. Following the introduction of $I$ individuals in the event, $q^{\star}$ of them stay at the origin node, and $I-q^{\star}$ walk to the next node. As the wave of walking individuals move along the line, $q^{\star}$ stay at every node until the number of remaining individuals reaching a node is below $q^{\star}$. We calculate first the radius of congestion $r_c$, defined as the distance to the closest non congested stop in each direction. Considering that the event individuals split in two streams --left and right--, and that part of the individuals will reach their destination without ever entering a vehicle,  $r_c(I)$ towards each direction can be obtained analytically (See SI section S1). In the limit of an infinite lattice ($L\rightarrow \infty$), the analytical solution for $r_c(I)$ yields
\begin{equation}\label{q}
r_c(I)\sim \frac{I}{2\, q^{\star}}.
\end{equation}
In Fig. \ref{Figure5}A, we compare the analytical expression obtained for $r_c(I)$ with numerical simulations finding a good agreement between them.

From $r_c(I)$, we can analytically derive the average delay $\Delta\tau_{av}(I)$. We need to consider the two types of individuals: those that stay in each of the queues waiting for their turn, and those who walk until their final destination. In the first case, the delay is the sum of the time until all the previous nodes empty plus the time in emptying the local queue. In the second case, the delay is the distance between the event and their destination  divided by the difference in speeds between walking and using a vehicle. The detailed analytical calculations, included in the SI section S1, display a good agreement with the simulations as depicted in Fig. \ref{Figure5}b. In the limit $L\rightarrow \infty$, our analytical solution approaches $\Delta\tau_{av}\sim \frac{f\,I}{4\,c}$. The scaling in Fig. \ref{Figure5}b applies to an event located at the center of the lattice, yet the analytical results hold for an event placed anywhere (SI Fig. S1).

\paragraph*{2D lattices.}

Empirical networks are closer to $2D$ lattices. The linear scaling obtained in 1D lattices is a product of the absence of alternatives: individuals must take the only existing line or walk to the next location. In higher dimensions, however, there is not only a degeneration of optimal paths (several isochronous paths lead to the same destination) but as they travel and find congested nodes many more alternatives become available. Moreover, those alternatives lead to a fragmentation of the flows in every intersection reaching faster a value under the vehicle capacity limit. 2D lattices of side $L$ are composed of $4 \,L$ layers containing one directed line 
and the additional walking layer. A sketch of the walking layer and two of the four 
directions is depicted in the inset of Fig. \ref{Figure5}b. To estimate the extension of the congested area provided by $r_c(I)$ the relevant quantity is again $q^{\star}$ (Eq. \eqref{qstar}). Since four lines (nodes) concur at every location and each node has an independent queue, more than $q^{\star}$ individuals can be waiting at the same location. Considering that movements are directed and individuals are not allowed to return to previously visited nodes, not all directions will be selected. In fact, in 2D lattices, there will be three types of nodes: the node of the event that has $4$ suitable directions, those in the main axes departing from it, which have $3$ directions and the rest with only $2$. For the sake of simplicity, we assume that the event occurs in the center of the lattice (or that the lattice is infinite) so that all directions are equivalent. 

Unlike 1D lattices, now the number of suitable queues sitting at a distance $r$, $N_q(r)$, scales as $N_q(r)\sim r^2$ .
Dividing the space in equivalent quadrants and considering only individuals with destination within each quadrant $I'=\frac{I}{4}$, the results in 1D can be extended to calculate analytically the radius of congestion $r_c$ (See SI section S2). In the limit of $L\rightarrow \infty$, it goes as
\begin{equation}
r_c \approx \sqrt{\frac{I'}{q^{\star} }} \sim \left( \frac{I}{4\,q^{\star}} \right)^{1/2}.
\end{equation}
While in the 1D lattice the radius of congestion was linear with $I$, in the 2D case it scales as $I^\frac{1}{2}$. The analytical results are confirmed by the simulations as depicted in Fig. \ref{Figure5}c. The average delay suffered by the individuals in the event can then be calculated separating again between those waiting at the queues and those reaching their destination by walking. The complete analytical derivation is shown in Section S2 of the SI and in the limit $L\rightarrow \infty$ it yields 
\begin{equation}
\Delta\tau_{av}=\frac{2\,f\,q^2\,I^{\frac{1}{2}}}{6\,c}.
\end{equation}
The results are displayed in Fig. \ref{Figure5}d, confirming the agreement between the analytical solution and the simulations. The dashed line, with an exponent $\frac{1}{2}$, serves as guide to the eye. 

\paragraph*{General dimensions.}

For a general dimension $D>2$, the scaling in the large $L$ and $I$ limits can be attained with an approximation. Assuming that $q^{\star}$ individuals remain at each node, we can estimate first the number of nodes congested as $Q_T \approx \frac{I}{q^{\star}}$. In the limit $L \to \infty$, they will occupy an hypervolume of the order of $\sim r_c^D$ so that the radius of the affected area can be approximated as
\begin{equation}
r_c \approx \left( \frac{I}{q^\star} \right)^{1/D} .
\end{equation}
If the total delay is dominated by the individuals waiting for a vehicle as occurs in lower dimensions and using the radius of congestion, $\Delta\tau_{tot}$ can be approximated as
\begin{equation} \label{rcongnd1}
\begin{aligned}
\Delta\tau_{tot} \approx \frac{{q^{\star}}^{2}f}{c} \sum_{r=0}^{r_c-1} N_q(r),
\end{aligned}
\end{equation}
where $N_q(r)$ is number of suitable queues that increases with the distance from the center of the lattice. The previous expression can be explained by the $q^{\star}$ individuals that stay at each node and the delay they incur $f\,q^{\star}/c$, which is the time that each node takes to empty. Considering that $\sum_{r=0}^{r_c-1} N_q(r) \sim {r_c}^{D+1} \sim {I}^{\frac{D+1}{D}}$ and dividing by $I$, we find that
\begin{equation}
\Delta \tau_{av} \sim I^{1/D}.
\end{equation}
In Fig. \ref{Figure5}e and f, we check by means of simulations that the scaling exponent of $1/D$ also holds in 3D lattices, showing how both $r_c$ and $\Delta \tau_{av}$ scale as $I^{\frac{1}{3}}$. All the results so far have been performed without background, yet both analytical calculations and simulations are extended to individuals participating in an event with background in the SI sections S1 and S2.

\begin{figure*}
\begin{center}
\includegraphics[width=\textwidth]{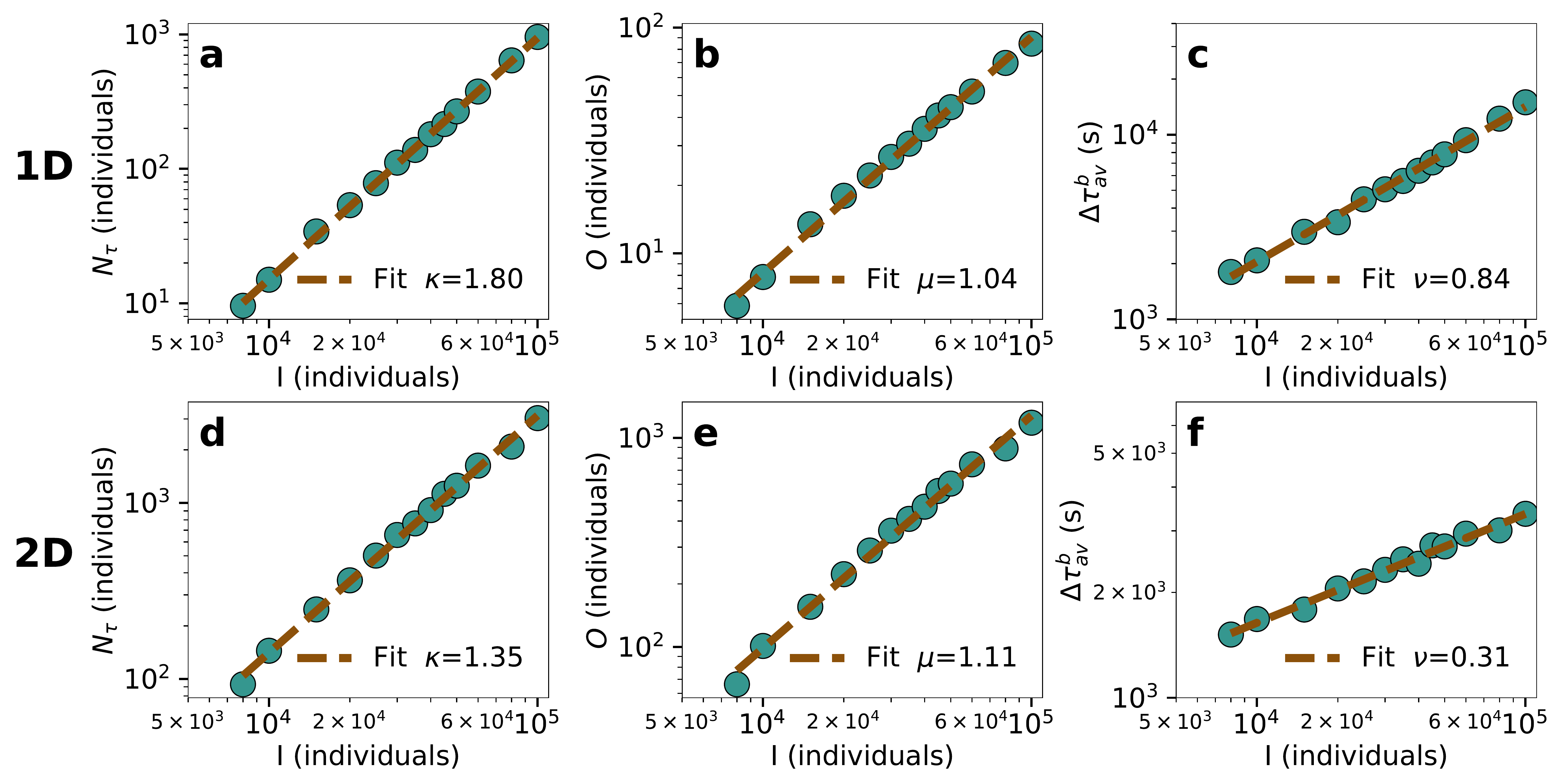}
\caption[Scaling for the background in regular lattices]{Scaling for the background in regular lattices, the parameters of the simulation are the same as in Fig. 4 of the main text. In a 1D lattice, scaling with $I$ of \textbf{a} the number of delayed individuals, \textbf{b} average delay per individual and \textbf{c} number of origins affected. In a 2D lattice, scaling with $I$ of \textbf{d} the number of delayed individuals, \textbf{e} average delay and \textbf{f} number of origins affected. In the case of the background, only individuals who waited for at least two vehicles were considered as delayed.
\label{Figure7}}
\end{center}
\end{figure*}

\subsubsection*{Background individuals}

Apart from the individuals involved in the event, other citizens performing their daily tasks may suffer the consequences of the congestion. There we study the effect of the event in the background individuals in the case of one and two dimensional lattices. To quantify it,  we use the same three metrics studied in cities: the number of delayed individuals $N_\tau$, the number of unique origins of the individuals delayed $O$ and their average delay $\Delta \tau^b_{av}$. In Figs. \ref{Figure7}a,b and c, we show the scaling of those three metrics with $I$ in 1D lattices. In Fig. \ref{Figure7}a, we have the scaling of the total delayed individuals $N_{\tau} \sim I^{\kappa}$, observing a super-linear scaling with an exponent close to $2$. In this case, we have not been able to find an exact analytical solution, yet the scaling exponent can be explained by a combination of two factors: (i) the origins of the potential individuals affected are the congested nodes, whose number ($Q_T$) scales linearly with I; (ii) the time needed by all the nodes to process the queue is proportional to $r_c$ and, thus, scales as $I$ in 1D (if we assume that all the nodes get congested when the event occurs). In Fig. \ref{Figure7}b, we see a linear scaling of the total number of origins affected $O \sim I^{\mu}$, which can be explained by the linear scaling of $r_c$. The effect is better visible for relatively high $\rho$. Finally, in Fig. \ref{Figure7}c, we show the scaling of the average delay per delayed individual in the background $\Delta\tau^{b}_{av} \sim I^{\mu}$. As we could expect, it has a similar exponent to the delay of the individuals participating in the event since they use the same routing protocol, the same network and the only difference lies on the origin of their trips.

We show the scaling of the background in 2D in Figs. \ref{Figure7}d, e and f. In the case of the delayed individuals (Fig. \ref{Figure7}d), the exponent can be explained by the same previous argument. In the two dimensional case, the scaling of the number of congested nodes is also linear with $I$, yet the time of the nodes to process the queue grows as $r_c$, which scales as $I^\frac{1}{2}$. In Fig. \ref{Figure7}e, the number of origins affected scales with $I$, similarly to 1D lattices, as a consequence of the linear growth of the number of congested nodes. Finally in Fig. \ref{Figure7}f, the average delay of the individuals in the background displays the same scaling as those in the event. The scaling observed in 2D lattices resembles the one obtained in empirical networks, evincing that the mechanisms behind are the same.

Our results in lattices in the context of cities indicate that most of the local structures of the public transportation network are almost two dimensional, and the observed deviations are consequences of the heterogeneous topology of real transportation network and the mixture of lines with uneven periods and capacities (see SI section S3, Figs. S4 and S5). The scaling above $0.5$ indicates that the local structure of the transportation network has less alternatives than 2D lattices or the destinations are biased towards one side. For example, areas containing only a terminal line typical fall in this category. Conversely, exponents below $0.5$ correspond to a local dimension greater than two as shown in the SI section S3, corresponding to areas with high capacity lines nearby.

\section*{Discussion}

We focus in this work on the effect of massive events on urban transportation systems. The agglomeration of individuals gives raise to congestion and produces delays in the trips of both individuals participating in the event and those doing their normal trips. We have introduced a model able to simulate the travels across public transportation system featured as multilayer networks. The origin and destination of the agents' trips are extracted from data in eight worldwide cities. In case of congestion, individuals can reroute their trajectories in an attempt to minimize travel times. We observe a scaling relation between the delay induced to individuals, as well as the number of congested nodes and affected trip origins, with the number of participants in the event $I$. 

To gain further insights, we have analyzed the case in which the transportation network is a regular lattice. This is a simplistic configuration, unrealistic out of $2D$, but it allows for analytical treatment. In fact, we find analytical solutions both in one and two dimensions and an approximate one for dimensions $D >2$. The average delay per delayed individual and the radius of the congested area both scale as $I^{1/D}$. The number of congested nodes, on the other hand, scales linearly with $I$ independently of the dimension. The individuals in the background are affected by the event too. The number of affected individuals, trip origins and average delay scale with $I$ with different exponents, which can be explained and are related to the lattice dimensionality. Overall, our analytical results in lattices shed light to the mechanisms governing the scaling in transportation networks of cities.

We simulated massive events in eight global cities, showing that a local dimension can explain their scaling. The heterogeneous topology and the multimodal nature of public transportation networks gives rise to a whole range of exponents between $0.3$ and $0.7$. These values imply that the local dimensionality of some locations in the cities approach 1D, for example terminal areas of the transportation lines where the individuals do not have many choices regarding mobility. On the other hand, there are areas with local dimension larger than $2$ as a reflection of the presence of high capacity lines nearby. Furthermore, we introduce an alternative way of measuring the local dimension by studying how the total transportation capacity grows with the distance to the event that show a good agreement with the simulation results. As for lattices, the background mobility also displays a scaling with $I$. Overall, our model allows us to determine the weakest (or strongest) points in a city for the organization of massive events by mapping the local dimension of the network.


\section*{Acknowledgements}

Partial financial support has been received from the Agencia Estatal de Investigacion (AEI) from Spain and Fondo Europeo de Desarrollo Regional FEDER/EU under Project PACCS-CSIC (RTI2018-093732-B-C22) and from the AEI, through the Maria de Maeztu Program for units of Excellence in R\&D (MDM-2017-0711). AB is funded by the Conselleria d'Educaci\'o, Cultura i Universitats of the Government of the Balearic Islands and the European Social Fund. ML thanks the French National Research Agency for its financial support (project NetCost, ANR-17-CE03-0003 grant).

\appendix

\setcounter{figure}{0}
\renewcommand\thefigure{A\arabic{figure}}
\renewcommand\theequation{A\arabic{equation}}

\section{1D  analytical results}


In this section, we provide the analytical solution for the scaling of the average delay $\Delta \tau_{av}$ with the number of individuals $I$ in the event for 1D regular lattices. First of all, and independently of the dimension, our system is mainly characterized by two speeds: $v_1$ for the vehicles and $v_2$ for the walking layer. The travel time between two nodes in each layer is then calculated by dividing the distance $\ell$ of a link by the corresponding speeds. Additionally, the transportation lines have two parameters: the capacity $c$, which is the total number of individuals that fit in a vehicle, and the period of the line $f$, which is the time elapsed between two consecutive vehicles. The variables of the system are thus:
\begin{equation}
\label{pformulation}
\begin{aligned}
v_1 \ vehicle\ speed ;\\
v_2 \ walking\ speed ;\\
c  \ capacity\ of \ the\ vehicles;\\
f	\ period \  of \ the\ line;\\
t_1=\frac{\ell}{v_1}; \\
t_2=\frac{\ell}{v_2} ;\\ . 
\end{aligned}
\end{equation}

We start by calculating the amount of individuals that fit in the queue of a node until walking to the next node becomes preferable. Recall that individuals use optimal paths to estimate travel times and to take routing decisions. In this case, the estimated time to empty a queue $q_i$ observed by individual $i$ is
\begin{equation} \label{eq11}
t_{wait,i}=\left(\frac{1}{2}+\left[\frac{q_i}{c}\right]\right) \, f. 
\end{equation}
The critical number of individuals that makes walking a better option $q^{\star}$ is obtained by matching the effective travel time considering congestion $t_1+t_{wait,i}$ and the walking time between two consecutive stops $t_2$:
\begin{equation}
t_2 = t_1+t_{wait,i}=\left(\frac{1}{2}+\left[\frac{q^{\star}}{c}\right]\right) \, f. 
\end{equation}
The expression for $q^{\star}$ is then given by
\begin{equation}\label{pformulation1}
q^{\star}=  \left[\frac{t_2-t_1}{f}+\frac{1}{2}\right] \, c .
\end{equation}
The value $q^{\star}$ captures the number of individuals that will wait at each node, and will be used to calculate the total number of nodes congested.

The individuals in the event are divided between left and right, with a rate that depends on the position of the origin in the lattice. For instance, if the event is located in an extreme, all individuals go towards one side, while if it is located in the middle, the flow splits in two.  The number of individuals that go towards each side $I_{right}$ and $I_{left}$ in a lattice of size $L$ and with an event position $x$, counting locations from $0$ to $L-1$, can be calculated as
\begin{equation}\label{pformulation2}
\begin{aligned}
I'=
\begin{cases}
I_{right}=\frac{I\, (L-1-x)}{L},\\
I_{left}=\frac{I \, x}{L} ,
\end{cases}
\end{aligned}
\end{equation}
which is a consequence of the uniform distribution of destinations along the space. Considering that each direction can be treated independently, from now on, $L'$ and $I'$ correspond to the length of the segment and the rate of individuals in the chosen direction ($L' = x$ or $L-1-x$, respectively). When the flow of individuals arrive at a node, $q^{\star}$ of them will stay and the rest of them will walk to the next lattice location. Assuming the conservation of individuals, we can write $I'$ in terms of the people who remain at a node waiting for a vehicle and those that walk all along to their destination. Defining the radius of congestion $r'_c$ as the number of congested nodes in each direction, this expression takes the form
\begin{align}
I' = & \, \overbrace{q^{\star}}^{\mbox{node }0}  + \overbrace{q^{\star} + \frac{I'-q^{\star}}{L'}}^{\mbox{node }1} + \overbrace{q^{\star} + \frac{I'-2\, q^{\star}}{L'} - \frac{I'-q^{\star}}{L'^2}}^{\mbox{node }2} \\  
& + \overbrace{q^{\star} + \frac{I' - 3\, q^{\star}}{L'} - \frac{2\, I'- 3\, q^{\star}}{L'^2} + \frac{I'-q^{\star}}{L'^3}}^{\mbox{node }3} + \ldots \nonumber \\
& \nonumber\\ 
 \approx & \sum_{\beta = 0}^{r'_c-1} \left[ q^{\star} + \sum_{\alpha = 1}^{\beta} (-1)^{\alpha+1} \frac{\binom{\beta-1}{\alpha-1} \, I' - \binom{\beta}{\alpha} \,q^{\star} }{L'^{\alpha}} \right]\nonumber,
\label{expan}
\end{align}
where the first $q^{\star}$ is the contribution of the queue of the origin node $x$, the second $q^{\star}$ is the queue of the first node in the chosen direction and the third term with $1/L'$ corresponds to the number of individuals who arrive by walking and have as destination node $x\pm 1$. The ensuing terms are the contributions of node $x\pm 2$ and $x\pm 3$, etc. The series expansion continues until $r'_c-1$, after which only $q^{\star}$ or less individuals remain to allocate. Grouping terms of the same order in $1/L'$, one can get the following expression for $I'$:
\begin{align}
I' = & \, r'_c \, q^{\star} + \frac{(r'_c-1) \, I' -  \left(r'_c \, (r'_c- 1)/2 \right) \, q^{\star}}{L'} \\
& - \frac{\left((r'_c-1) \, (r'_c- 2)/2 \right) \, I' - \left(r'_c \, (r'_c- 1) \, (r'_c-2)/6 \right) \, q^{\star}}{{L'}^2} \notag \\
& +\ldots  \nonumber
\label{q_exp}
\end{align}
The coefficients of each term in $I'$ and $q^{\star}$ become more complicated as the order in $1/L'$ increases. Still, it is possible to find a closed form for the expansion:
\begin{equation}
I' \approx  r'_c \, q^{\star} + \sum_{\alpha = 1}^{r'_c-1} (-1)^{\alpha-1}\,  \frac{\binom{r'_c-1}{\alpha} \, I' - \binom{r'_c}{\alpha+1} \, q^{\star} }{L'^{\alpha}}  ,
\end{equation}
where the symbols $(.)$ are binomial coefficients. Note that we have written the expression as an approximate formula, because after node $r'_c$, there is a number of individuals (less than $q^{\star}$) who continue traveling and are not counted in this expression. Using the polynomial expansion and recalling that 
\begin{equation}
(1-1/L')^{r'_c-1} = \sum_{i = 0}^{r'_c-1} (-1)^i \, \binom{r'_c-1}{i} \, \frac{1}{L'^i},  
\end{equation}
we can rewrite the sum on the right as
\begin{equation}
I' \approx \left[ 1 - \left(1-\frac{1}{L'}\right)^{r'_c-1}     \right] \, I' 
    + \left[1 - \left(1-\frac{1}{L'}\right)^{r'_c} \right] \, q^{\star} \, L' .
\label{full_eq}
\end{equation}
This equation can be solved for $r'_c$ as a function of $L'$ and $I'$ yielding
\begin{equation}
r'_c \approx 1+ \frac{\ln \left(\frac{q^{\star} \, L'}{I'+ q^{\star} \, (L'-1)}\right)}{\ln\left(1-\frac{1}{L'}\right)} .
\label{eq_d}
\end{equation}
The total number of congested nodes will be then given by the sum of the radius of congestion towards both directions $Q_{T}= r_c(\mbox{Left})+ r_c(\mbox{Right})$. In the limit of $L' \to \infty$, the scaling of $Q_{T}$ with $I$ yields
\begin{equation}
Q_{T} \sim \frac{I}{q^{\star}},
\end{equation}
This linear dependency is maintained as long as $Q_{T} \ll L'$. Otherwise, when $Q_{T} \to L'$, the whole network is congested, $Q_T$ and the delay saturate with most agents walking to their destination. 

From the expression of $r_c'$, we can  estimate the average delay per individual $\Delta \tau_{av}$. First of all, we define the delay as the difference between the real and the expected travel time determined by the optimal path $\Delta \tau=\tau_{real}-\tau_{op}$. In our framework, individuals can suffer the delay in two different ways: either they walk all along until their final destination, or they stay in the queues and wait for their turn to enter into the vehicles. 
If they walk all the way, the total delay is calculated as the speed difference between walking and using a vehicle multiplied by the number of individuals. For each of the directions, the individuals walking are approximated by the terms displaying powers of $1/L'$ in the expansion of Eq. \eqref{expan}. By walking $i$ locations at speed $v_2$, the individuals incur in a delay of $i \, \ell/\Delta v$ (where  $\Delta v = v_1-v_2$). We can write the total delay of walkers as
\begin{align}
\Delta \tau_{tot, w} \approx & \,  \frac{\ell}{\Delta v} \, \Bigg \{ \overbrace{ \frac{I'-q^{\star}}{L'}}^{\mbox{node }1} + \overbrace{2\, \frac{I'-2\, q^{\star}}{L'} - 2\, \frac{I'-q^{\star}}{L'^2}}^{\mbox{node }2} \\  
&  + \overbrace{3\,  \frac{I' - 3\, q^{\star}}{L'} - 3\, \frac{2\, I'- 3\, q^{\star}}{L'^2} + 3\, \frac{I'-q^{\star}}{L'^3}}^{\mbox{node }3} \nonumber \\
& + \ldots \Bigg  \} . \nonumber
\label{expan_tau}
\end{align}    
Each term appears multiplied by the position of the node generating it. This expression can be compacted by grouping the terms of the same order in $1/L'$ to obtain
\begin{align}
\Delta \tau_{tot, w} \approx &  \frac{\ell}{\Delta v} \, \sum_{\alpha = 1}^{r'_c-1} \frac{(-1)^{\alpha+1}}{L'^{\alpha}} \, \left[ \frac{(r'_c- \alpha)\, r'_c}{(1+\alpha)} \, \binom{r'_c-1}{\alpha-1} \, I' \right. \nonumber\\
& +   \left[ - \frac{(r'_c-\alpha)\, ((\alpha+1)\,r'_c-1)}{(1+\alpha)\, (2+\alpha)} \, \binom{r'_c}{\alpha} \, q^{\star} \right] .
\end{align}
Suming the terms before $I'$ and $q^{\star}$ yields
\begin{align}
\label{tauw} 
\Delta \tau_{tot, w} \approx & \frac{\ell}{\Delta v} \, \left\{  \left[  L' - (L'+r'_c-1) \, \left(1-\frac{1}{L'} \right)^{r'_c-1} \right] \, I'   \right. \nonumber\\
& +   \left[  \left[ \frac{r'_c\, (1-r'_c)}{2} + L' \, (L'-1)  \nonumber \right.  \right. \\
& - \left. \left. L'\, (L'+r'_c-1) \, \left( 1- \frac{1}{L'}\right)^{r'_c} \right] \, q^{\star} \right\} .
\end{align}
This equation is valid for each of the directions and the total delay is, thus, the sum of $\Delta \tau_{tot, w} = \Delta \tau_{tot, w} (\mbox{Right})+ \Delta \tau_{tot, w} (\mbox{Left})$.  While it is important to have an accurate knowledge of the functional form of the delay,  the expression for $\Delta \tau_{tot, w}$ in Eq. \eqref{tauw} tends to zero when $L' \to \infty$ and, therefore, it does not contribute to the scaling of the delay.

The second contribution to the delay corresponds to individuals waiting at queues. The delay suffered by an individual depends on his/her position in the queue. Dividing the $q^{\star}$ individuals in packs of $c$, the total delay cumulated in a single node $\Delta \tau_{sn}$ is given by
\begin{equation}
\Delta \tau_{sn} =  c\, f\, \sum_{i=1}^{p-1} i + (q^{\star}- p\, c)\,p \, f ,
\end{equation}
where $p = [q^{\star}/c]$ is the integer part of the ratio. The sum runs over all the packs except the first, which has no delay, and the last, which does not necessarily have $c$ individuals. Solving the sum, we get
\begin{equation}
\Delta \tau_{sn} = f \, p \, \left( \frac{c\, (p-1)}{2} + (q^{\star}- p\, c) \right).
\label{tau_sn}
\end{equation}
For the $q^{\star}$ individuals remaining at each of the queues, we have to add a delay of $f\, q^{\star}/c$ per link walked. This gives us the total delay of
\begin{align}
\label{totaldelay}
\Delta \tau_{tot,v} \approx & \, \overbrace{ f \, p \, \left( \frac{c\, (p-1)}{2} + (q^{\star}- p\, c) \right) }^{\mbox{node }0}+
 \nonumber\\
&  \overbrace{ \frac{{q^{\star}}^2}{c} \, f +  f \, p \, \left( \frac{c\, (p-1)}{2} + (q^{\star}- p\, c) \right)}^{\mbox{node }1} \nonumber \\  
& + \overbrace{2\, \frac{{q^{\star}}^2}{c} \, f + f \, p \,\left( \frac{c\, (p-1)}{2} + (q^{\star}- p\, c) \right) }^{\mbox{node }2} + \ldots ,
\end{align}    
The previous expression can be reordered to obtain
\begin{equation}
\Delta \tau_{tot,v} \approx r'_c\, f \, p \, \left( \frac{c\, (p-1)}{2} + (q^{\star}- p\, c) \right) + \frac{{q^{\star}}^2\, f\, r'_c\, (r'_c-1)}{2\, c}.
\label{tau_v}
\end{equation}
As before, this is the delay in one direction. Therefore, to calculate the average delay per individual $\Delta \tau_{av}$ we need to sum $\Delta \tau_{tot,v}$ in both directions, add it to the one of the walkers $\Delta \tau_{tot,w}$ and divide by the total number of individuals in the event $I$, yielding
\begin{align}
\Delta \tau_{av} = &  \left\{ \Delta \tau_{tot,v} (\mbox{Right}) +\Delta \tau_{tot,v} (\mbox{Left}) \right. \\ & \left. + \Delta \tau_{tot,w} (\mbox{Right}) +\Delta \tau_{tot,w} (\mbox{Left}) \right\}/I . \notag
\end{align}
Considering that in the limit $L \to \infty$, $r'_c \sim I'$, $\Delta \tau_{tot,w} \to 0$ and, according to Eq. \eqref{tau_v}, $\Delta \tau_{tot,v} \sim I^2$, the scaling of  $\Delta \tau_{av}$ with $I$ yields
\begin{equation}
\Delta \tau_{av} \sim \frac{f\, I}{2\, c\,}. 
\end{equation}

\begin{figure}[ht]
\centering
\includegraphics[width=8cm]{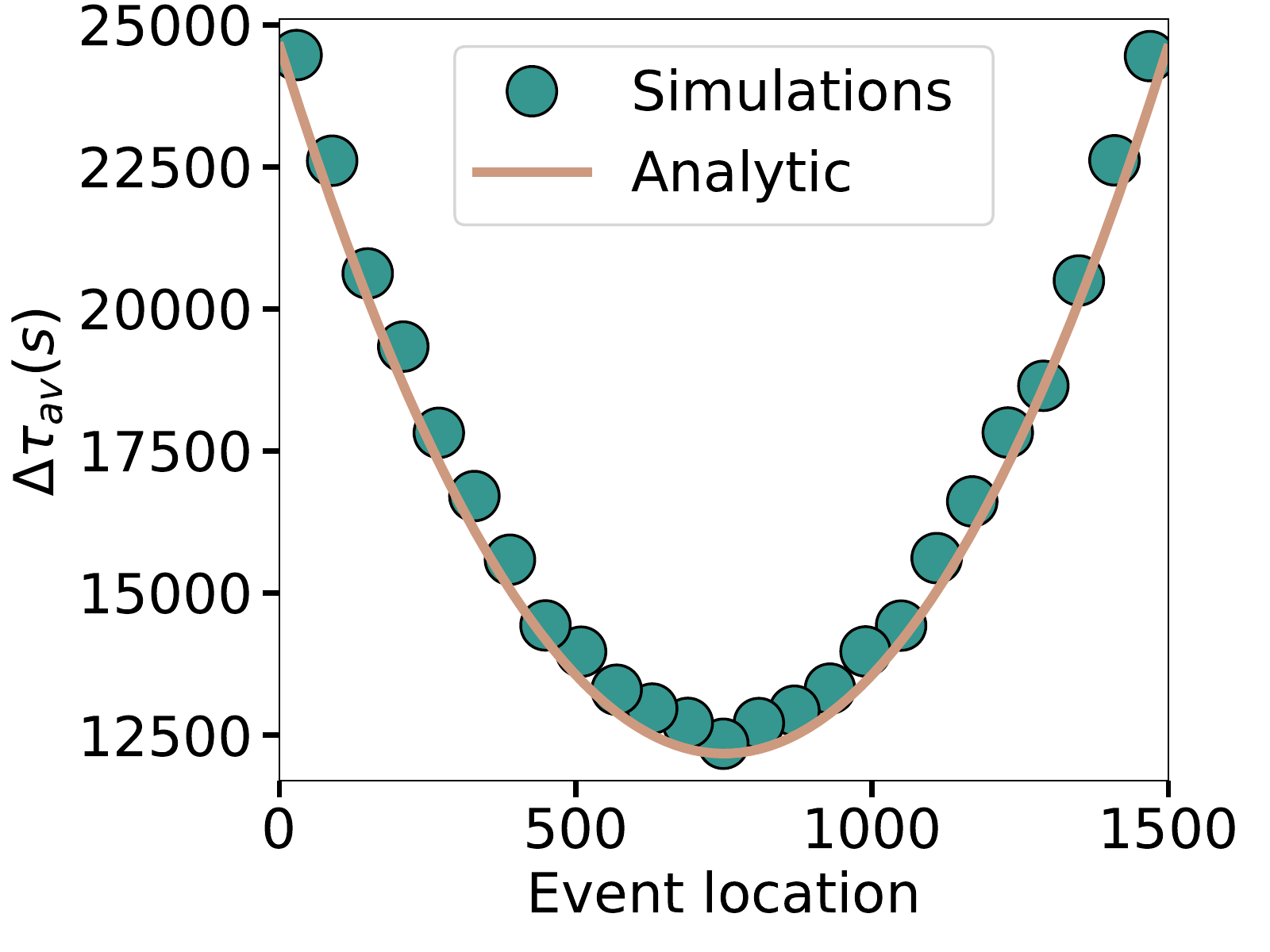}
\caption[Average delay as a function of the event location]{Average delay per individual and its analytical prediction in an 1D lattice with the same event $I$ placed in different lattice locations.}
\label{Figure8}
\end{figure}

In finite networks, we can also obtain the number of congested nodes and average delay for any event location. In Fig. \ref{Figure8}, we show $\Delta \tau_{av}$ as a function of the position $x$. Due to the split of the stream of individuals, the lowest delay appears for an event in the center of the lattice. Conversely, the highest congestion occurs when the event is introduced in the lattice extremes.

\begin{figure}[ht]
\centering
\includegraphics[width=7cm]{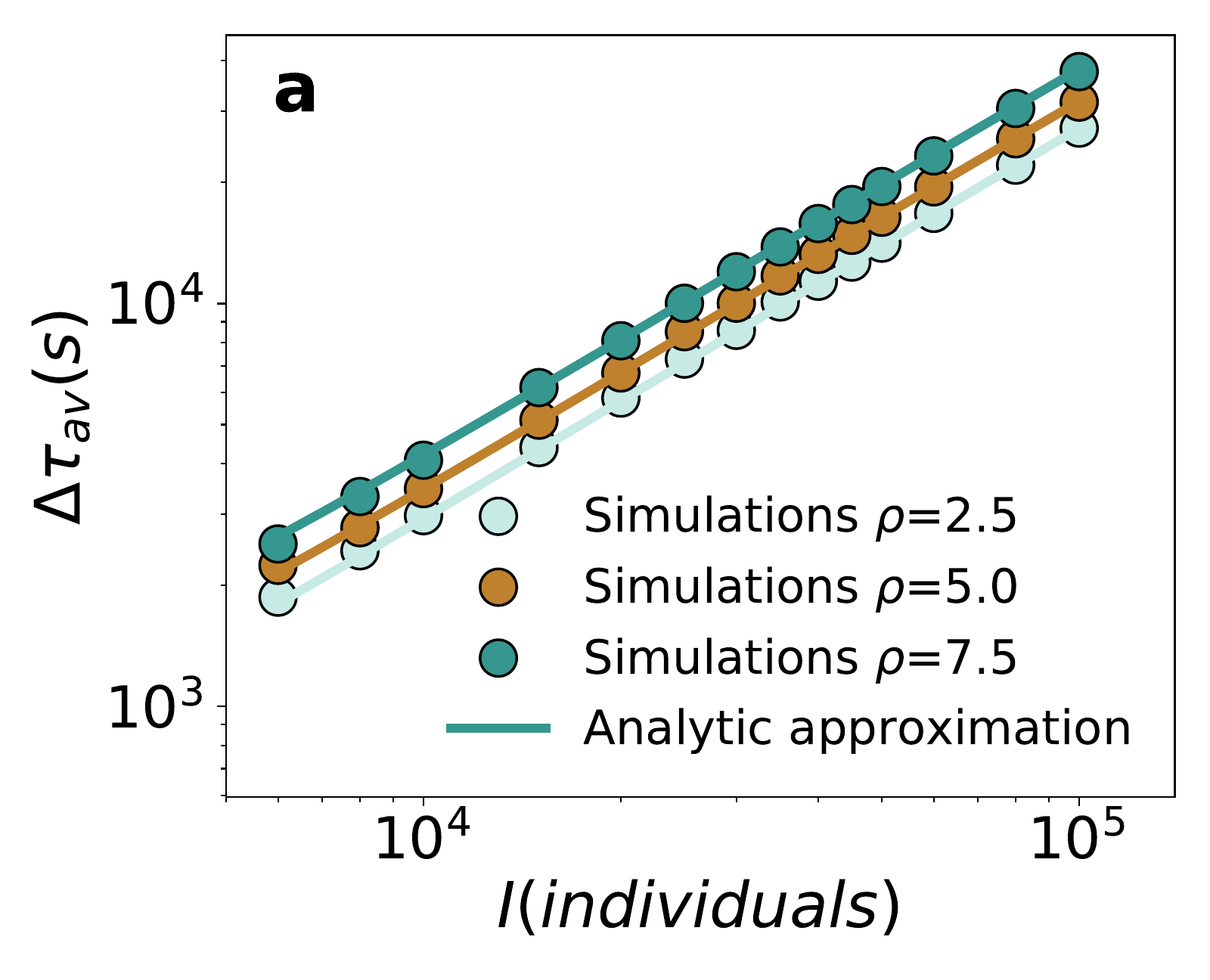}
\includegraphics[width=7cm]{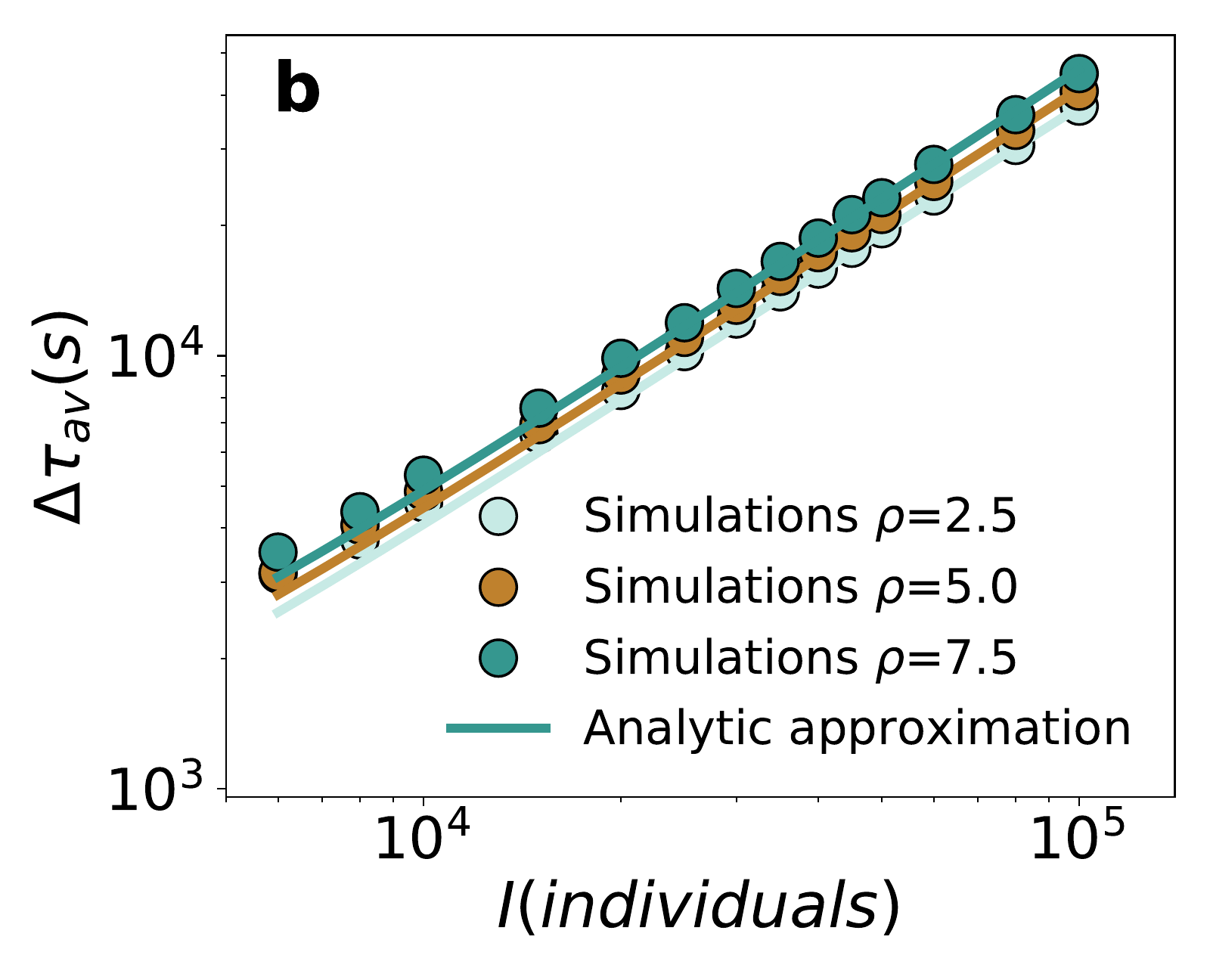}
\caption[Scaling of the average delay with $I$ in a 1D lattice with background]{For the 1D lattice with background,  scaling of the average delay of agents depending on the individuals introduced in the event. On the left the event is placed in the extreme of the lattice, while on the right it is at the center.}
\label{Figure9}
\end{figure}

These previous results can be extended to take into account the effect of background individuals. If the background is active, a number $\rho$ of trips will be generated at each time step with an origin and destination selected at random. While conserving the scaling, the background increases the average delay because the new passengers reduce the effective vehicle capacity. We need now to separate the contributions of the individual flows to the left and to the right. The effective capacity for locations to the left of the event can be approximated as
\begin{equation}
\label{cf1}
c_{eff}^l(x) = c - \rho \,  f \, \frac{x\, (L-x)}{L \, (L-1)} ,
\end{equation}
while for those to the right $ c_{eff}$ is
\begin{equation}
c_{eff}^r(x) = c - \rho \,  f \, \frac{(x+1)\, (L-1-x)}{L \, (L-1)} .
\label{cf2}
\end{equation}
Here $x$ refers to the node location, the factors $x\, (L-x)/L \, (L-1)$ and  $(x+1)\, (L-1-x)/L \, (L-1)$ are the ratio of shortest paths that go from $x$ to the left and to the right, respectively. They related to  the betweenness of  $x$. Eqs. \eqref{cf1} and \eqref{cf2} show that the impact of the background is maximum in the central node, while it decreases for the nodes at the extremes. The location of the special event is, therefore, of great relevance since nodes no longer have equal capacity.

To obtain the new expression for the total delay with background, we need to repeat the calculations of Eq. \eqref{totaldelay} using $c_{eff}'(x)$ (the one corresponding to the given direction) instead of $c$. To be more specific, if $x$ is the event location, the $i-th$ congested node to the right and to the left will have an effective capacity $c_{eff}^r(x+i)$ and $c_{eff}^l(x-i)$, respectively.  The sum of the terms of Eq. \eqref{totaldelay} in each direction can then be written as
\begin{align}
\Delta \tau_{tot,v}  = & \sum_{i=0}^{r'_c}  f\, p(i)\left( \frac{c\, (p(i)-1)}{2}+(q^{\star}-p(i)\, c)\right)  + \nonumber \\  & \sum_{i=1}^{r'_c-1} i\, \frac{f\, {q^{\star}}^2}{c_{eff}'(i)} ,
\label{totaldelayceff}
\end{align}
where $p(i)$ is the integer part of $[q{^\star}/c_{eff}'(x \pm i)]$ with the sign and the expression for $c_{eff}$ depending on the direction over the lattice. The integer division in the first summation makes obtaining an analytical expression very difficult. There is, however, an extra approximation that we can take to get an analytical solution for $\Delta \tau_{av}$. This passes through substituting $c_{eff}'(x\pm i)$ by an average value $c_{eff}'(x\pm {r'_c}/2)$. If we introduce this approximation of $c_{eff}'$ into Eq. \eqref{totaldelayceff}, we obtain for each direction 
\begin{align}
\Delta \tau_{tot,v}  = \,  & r'_c\, f \, p \,  \left[ \frac{(p-1) \, c_{eff}'(x \pm \frac{r'_c}{2})}{2}  
+  q^{\star} - p\, c_{eff}'(x \pm \frac{r'_c}{2}) \right]   \nonumber \\
& + \frac{q^{\star}\, f\, r'_c\, (r'_c-1)}{2\, c_{eff}'(x \pm \frac{r'_c}{2})} ,
\label{totaldelayceff1}
\end{align}
where $p = [q{^\star}/c_{eff}'(x \pm r'_c/2)]$. As before, the average delay $\Delta \tau_{av}$ is obtained summing the total delay in both directions and dividing by $I$.  As depicted in Fig. \ref{Figure9}, there is a good agreement between the simulations and the analytic approximation.  The average delay per individual increases with the intensity of the background $\rho$ due to the reduction of the effective capacity . Moreover, the impact of the background is more pronounced in the center of the lattice than at the extremes. We also corroborate that the background does not affect the scaling.

\section{ 2D  analytical results}

\setcounter{equation}{25}

In this section, we provide the analytical solution for the scaling in 2D lattices. As detailed in the main manuscript, in two dimensions there are three types of nodes according to the number of suitable directions, and the number of queues at distance $r$ is given by
\begin{equation}
\label{qr}
N_q(r) = 4 \, \delta_{r,0} + 12 \, H_{r,1} + 8 \, (r-1) \, H_{r,2} .
\end{equation}
where $\delta_{r,0}$ is the Kronecker delta and $H_{r,1}$ is the discrete step function. 
By dividing the space in four equivalent quadrants and considering only the individuals $I'$ whose destination is within each quadrant, we can write $I'$ as a function of the individuals that stay at a queue, walk all along to their destination and the distance to the closest non congested location $r'_c$ as 
\begin{align}
\label{expan2d}
I' \approx & \, \overbrace{q^{\star}}^{\mbox{distance }0}  + \overbrace{ 3 \, q^{\star} + \frac{I'-q^{\star}}{L_x' \, L_y'}}^{\mbox{distance }1} + \nonumber \\
& \overbrace{5 \, q^{\star} + \frac{I'- 4\, q^{\star}}{L_x' \, L_y'} - \frac{I'-q^{\star}}{(L_x' \, L_y')^2}}^{\mbox{distance }2} \\  
& + \overbrace{7 \, q^{\star} + \frac{I' - 9\, q^{\star}}{L_x' \, L_y'} - \frac{2\, I'- 5\, q^{\star}}{(L_x' \, L_y')^2} + \frac{I'-q^{\star}}{(L_x' \, L_y')^3}}^{\mbox{distance }3} + \ldots \nonumber \\
& \nonumber\\ 
 = & \sum_{\beta = 0}^{r_c-1} \Bigg \{    (2\, \beta+1)\,  q^{\star} \nonumber \\  &  +  \sum_{\alpha = 1}^{\beta} (-1)^{\alpha+1} \,  \frac{\binom{\beta-1}{\alpha-1} \, I' - \frac{2\,\beta - (\alpha-1) }{(\alpha+1)}\, \binom{\beta}{\alpha} \,q^{\star} }{(L_x' \, L_y')^{\alpha}} \Bigg \} \nonumber,
\end{align}
where the lateral size of the lattice is $L$ and $(x,y)$ is the position of the event in the network (allegedly, $(L/2,L/2)$). $L_x'$ ($L_y'$) is defined depending on the quadrant and the direction of the individuals that are leaving the location of the event. If the movement is towards the right (up), then it is defined as $L_x' = L - x-1$ ($L_y' = L - y-1$), otherwise, if it is to the left (down) it is defined as $ L_x' = x$ ($L_y' = y$).
Likewise 1D lattices, we sum the coefficients appearing in terms of equal power in $(L_x' \, L_y')$ in Eq. \eqref{expan2d} to obtain
\begin{align}
\label{expan2d2}
I' \approx & \, {r'_c}^2 \, q^{\star} + \sum_{\alpha = 1}^{r'_c-1} (-1)^{\alpha+1} \frac{\binom{r'_c-1}{\alpha} \, I' - \frac{2\, r'_c-\alpha}{(\alpha+2)}\, \binom{r'_c}{\alpha+1} \,q^{\star}}{(L_x' \, L_y')^{\alpha}} .
\end{align}
Using the expansion of the binomial $(1-1/(L_x'\,L_y'))^\beta$, this expression can be transformed into
\begin{align}
\label{expan2d3}
I' \approx & \, \left[1-\left(1-\frac{1}{(L_x'\,L_y')}\right)^{r'_c-1}\right] \, I' \\
& + q^{\star}\, \left[ (L_x'\,L_y')\,  \left(2\, r'_c+1-\left( 1-\frac{1}{(L_x'\,L_y')}\right)^{r'_c} \right) \right. \nonumber\\
& \left. - 2\, (L_x'\,L_y')^2 \,   \left(1-\left( 1-\frac{1}{(L_x'\,L_y')}\right)^{r'_c} \right) \right] \nonumber . 
\end{align}
Solving the equation in $r'_c$, we obtain
\begin{align}
\label{rc2d}
& r'_c =   - \frac{1}{2} +   L'  \nonumber \\ & - \frac{Plog\left(   (1-\frac{1}{L'})^{L'- \frac{3}{2}} \, (\frac{q^{\star}  -I'}{2\, L'\, q^{\star} } + L'-\frac{3}{2}) \, \log(1-\frac{1}{L'})    \right)}{\log(1-\frac{1}{L'})} \nonumber ,
\end{align}
where $Plog()$ is the product log function and $L' = L_x' \, L_y'$. Asymptotically, in the limit of large lattices $L' \to \infty$, this expression goes with $I'$ (and with $I$) as
\begin{equation}
r'_c \approx \sqrt{\frac{I'}{q^{\star} }} \sim \left( \frac{I}{4\, q^{\star}} \right)^{1/2}.
\end{equation}

From $r'_c$, we can calculate the average delay suffered by the individuals participating in the event. The most straightforward expression is the one concerning the delay of people who arrive at the final destination by walking. These are individuals whose trip destination fall within the congested area, and given the long queues, find more profitable to walk all the way. Their numbers correspond to the terms with the powers in $1/(L_x'\, L_y')$ in Eq. \eqref{expan2d} and their delay depends on the difference in speed between the vehicles and walking. Recovering the Eq. \eqref{expan_tau} for 1D and adapting it to 2D, we get
\begin{widetext}
\begin{align}
\label{tauw_2d}
\Delta \tau_{tot,w} = & \,  \frac{4\, \ell}{\Delta v} \, \{ \overbrace{  \frac{I'-q^{\star}}{L_x' \, L_y'}}^{\mbox{distance }1}  
+\overbrace{2 \, ( \frac{I'- 4\, q^{\star}}{L_x' \, L_y'} - \frac{I'-q^{\star}}{(L_x' \, L_y')^2})}^{\mbox{distance }2}  + \overbrace{3 \, ( \frac{I' - 9\, q^{\star}}{L_x' \, L_y'} - \frac{2\, I'- 5\, q^{\star}}{(L_x' \, L_y')^2} + \frac{I'-q^{\star}}{(L_x' \, L_y')^3})}^{\mbox{distance }3} + \ldots  \nonumber   \\
\nonumber\\ 
 =  &  \,  \frac{4\, \ell}{\Delta v} \,   \sum_{\alpha = 1}^{r_c-1}  \frac{(-1)^{\alpha+1}}{(L_x' \, L_y')^{\alpha}} \,  \left\{  \frac{(r_c-\alpha)\, r_c}{1+\alpha} \, \binom{r_c-1}{\alpha-1} \, I'  \frac{(r_c-\alpha) \, ((2\, \alpha+4)\, \ r_c^2 -(\alpha^2+2\, \alpha+3)\, r_c+\alpha-1)}{(\alpha+1)\, (\alpha+2)\, (\alpha+3)}\,  \binom{r_c}{\alpha} \,q^{\star} \right\}. 
\end{align}
\end{widetext}
Summing each term of $I$ and $q^{\star}$, the result is 
\begin{widetext}
\begin{align}
\label{tauw_2d2}
&\Delta \tau_{tot,w} =  \frac{4\, \ell}{\Delta v} \, \left\{ \left[ L' - (L'+r_c-1) \, x^{r_c-1} \right] \, I'
 +\left[   
(4\,  L' (6 \, (-1 + x^{r_c}) \, L'^3\,  (12 - 5 \, r_c + 12 \, r_c^2 + 2\,  r_c^3) \right. \right. \\
& \nonumber \\
&  +  6 \, L'^2 \, (18 - 18 \, x^{r_c}  
+ 3 \, (3 + x^{r_c})\, r_c - 
         5 \, (-2 + 3 \, x^{r_c}) \, r_c^2 + (3 + 9 \, x^{r_c}) \, r_c^3 +  2\, x^{r_c} r^4) \nonumber \\
 & \nonumber \\
&   -   3\, L' \, (12 - 12 x^{r_c} - 
         4 \, (-7 + x^{r_c}) \, r_c  
        + (17 - 
            6\,  x^{r_c})\,  r_c^2 + (-9 + 22\,  x^{r_c}) \, r_c^3 + 
         2 (-5 + x^{r_c}) \, r_c^4 - 2\,  r_c^5)\nonumber \\
 & \nonumber \\
&  
     - 
      r_c (-6 - 29\,  r _c - 15\, r_c^2 + (25 + 6\,  x^{r_c})\,  r_c^3 + 21 \, r_c^4 + 
         4 \, r_c^5)) \nonumber \\
 & \nonumber \\
&       
         - 
   r_c^2 \, (6 + 11 \, r _c+ 6 \, r_c^2 + r_c^3) \, _pF_q(\{2, 2, 2, 
      1 - r_c\};\{1, 1, 5\};1/L'])
       \left. \left.   
      /(24 \, L' \, (1 + r_c) \, (2 + r_c) \, (3 + r_c)) \right ]\, q^{\star} \right\} \nonumber ,
\end{align}
\end{widetext}
where $x = (1 -1/L')$, $_pF_q()$ is the generalized hypergeometric function and $L' = L_x' \, L_y'$. Since this expression is a result of summing a series in terms of $1/L'$, and given that $r_c<<L$ its contribution tends to zero in the infinite size limit.

\begin{figure}[ht]
\centering
\includegraphics[width=8cm]{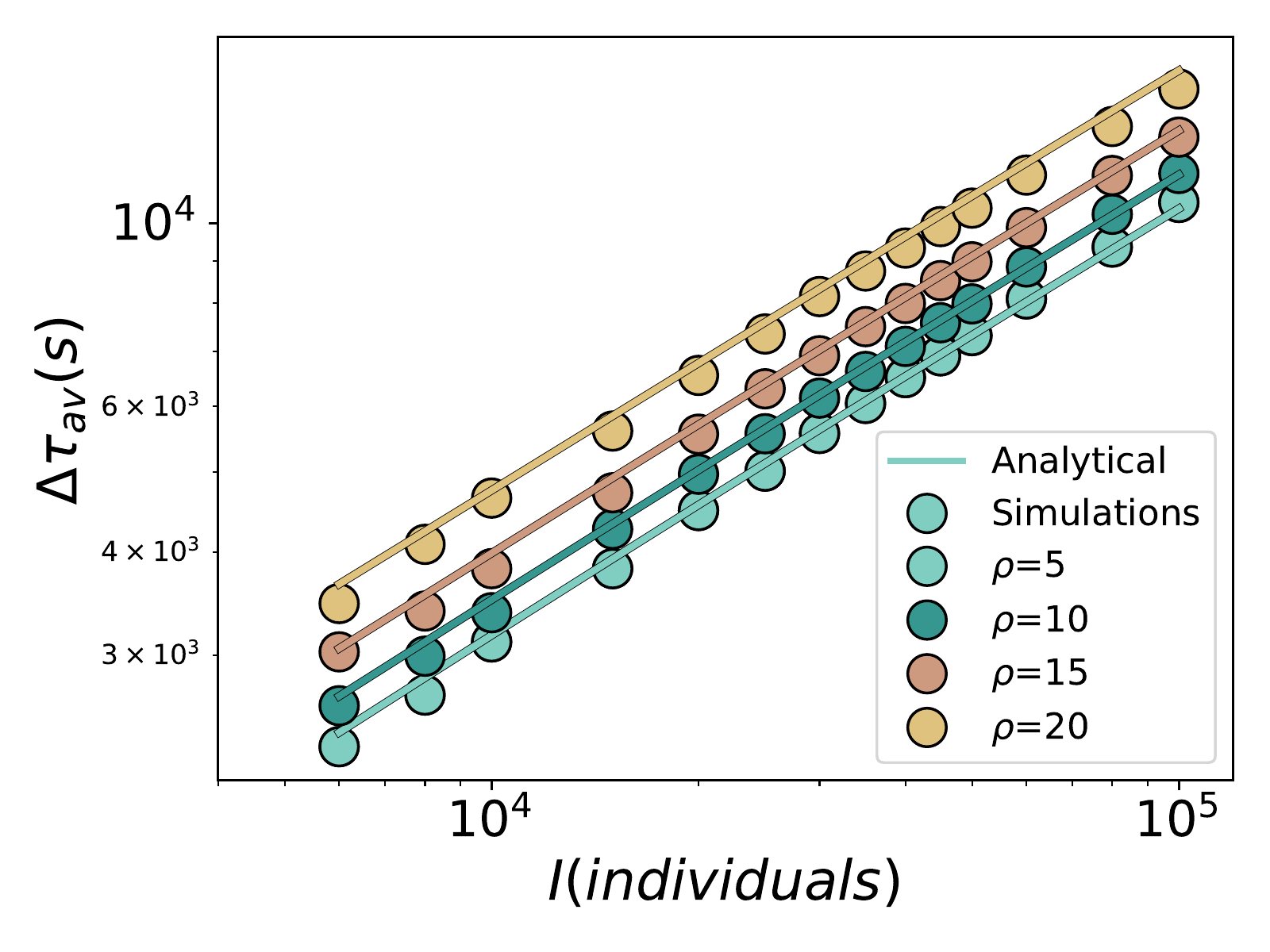}
\caption[Scaling of the average delay with $I$ in a 2D lattice with background]{Scaling of the average delay per individual with $I$ in a 2D lattice with background.}
\label{Figure10}
\end{figure}

As in any dimension, the most important contribution to the delay comes from the individuals using the vehicles. The calculations are similar to those of 1D, with the additional consideration that the locations have a different number of queues. Given that the delay accumulated while emptying is independent of the dimension (Eq. \eqref{tau_sn}), and taking as basis Eq. \eqref{totaldelay}, the total delay of the individuals using a vehicle in a 2D lattice can be written as  
\begin{align}
\label{totaldelay2d}
\Delta \tau_{tot,v} = & \, \overbrace{ 4\, f \, p \, \left( \frac{c\, (p-1)}{2} + (q^{\star}- p\, c) \right) }^{\mbox{distance }0} \nonumber \\  &  + \overbrace{ 12 \, \frac{{q^{\star}}^2}{c} \, f + 12\,  f \, p \, \left( \frac{c\, (p-1)}{2} + (q^{\star}- p\, c) \right)}^{\mbox{distance }1} \nonumber \\  
& + \overbrace{40\, \frac{{q^{\star}}^2}{c} \, f + 20\,  f \, p \,\left( \frac{c\, (p-1)}{2} + (q^{\star}- p\, c) \right) }^{\mbox{distance }2} + \ldots , 
\end{align}   
where to find the pre-factors for each term one must recall the expression for $N_q(r)$ in Eq. \eqref{qr}. Again, $p$ is the integer part of $q^{\star}/c$. Grouping together the different terms of Eq. \eqref{totaldelay2d}, one can obtain the expression
\begin{align}
\label{totaldelay2d_2}
\Delta \tau_{tot,v} = & \, \frac{{q^{\star}}^2}{c} \, f \, \sum_{r = 1}^{r_c-1} r\, (8 \, r +4) \nonumber \\ & + f \, p \, \left( \frac{c\, (p-1)}{2} + (q^{\star}- p\, c) \right) \, \sum_{r = 0}^{r_c-1} (8 \, r +4) .
\end{align} 
Finally, summing up the series with the number of queues yields
\begin{align}
\label{totaldelay2d_3}
\Delta \tau_{tot,v} = & \, \frac{{q^{\star}}^2}{c} \, f \, \frac{2\, (r_c-1)\, r_c\, (1+4\, r_c)}{3} \nonumber \\  &  + f \, p \, \left( \frac{c\, (p-1)}{2} + (q^{\star}- p\, c) \right) \, 4\, r_c^2 .
\end{align} 

The asymptotic behavior of $\Delta \tau_{tot,v}$ is obtained by plugging $r_c \sim I^{1/2}$ in Eq. \eqref{totaldelay2d_3}, and yields
\begin{equation}
\Delta \tau_{tot,v} \sim \frac{{q^{\star}}^2 \, f}{c} \, I^{3/2} .
\end{equation}
This means, dividing by $I$, that the average delay per delayed individual scales as 
\begin{equation}
\Delta \tau_{av} \sim ({q^{\star}}^2 \, f/c)\, I^{1/2} .
\end{equation}

The analytical solution outlined above, as well as the results shown in the main text, hold only when there are no individuals in the background. As in 1D lattices, the introduction of individuals in the background, modifies effectively the capacity of the lines and we need to find an expression for $c_{eff}(x,y)$. Luckily, the penalty for changing line included in our model simplifies the calculation of the betweenness centrality since there is a maximum of two path alternatives between any pair of nodes. In our framework, the edge betweenness centrality will depend on its location and direction. Therefore, as a function of the lattice side $L$ and the coordinates of the source node $(x_0,y_0)$ and target node $(x_1,y_1)$ of a given edge, can be written as:

\begin{widetext}
\begin{align}
\label{bet2d}
& b((x_0,y_0),(x_1,y_1))=  \,  \frac{\delta_{(y_1-y_0),-1}}{L^2\, (L^2-1)} \left[ (L-y_0)\, y_0+ (L-y_0)\ y_0\, x_0+(L-y_0)\, y_0\, (L-(x_0+1))  \right] \notag \\
& \\
& + \frac{\delta_{(y_1-y_0),1}}{L^2\, (L^2-1)}\left[ (y_0+1)\, (L-(y_0+1))+ (y_0+1)\, (L-(y_0+1))\, (L-(x_0+1)) +\right. \notag\\
& \nonumber \\
& \left. (y_0+1)\, (L-(y_0+1))\, x_0 \right] + \frac{\delta_{(x_1-x_0),-1}}{L^2\, (L^2-1)} \left[ (L-x_0)\, x_0+ (L-x_0)\ x_0\, y_0+\right. \notag\\
& \nonumber \\
& \left.+(L-x_0)\, x_0\, (L-(y_0+1))  \right] + \frac{\delta_{(x_1-x_0),1}}{L^2\, (L^2-1)}\left[ (x_0+1)\, (L-(x_0+1))+
\right. \notag\\
& \nonumber \\
& \left.+ (x_0+1)\, (L-(x_0+1))\, (L-(y_0+1)) +(x_0+1)\, (L-(x_0+1))\, y_0  \right]. \notag
\end{align}
\end{widetext}


The average number of individuals of the background going through a link can then be calculated as $\rho \,  b((x_0,y_0),(x_1,y_1))\, f$. As an approximation, we will use $c-\rho \,  b((\frac{L+r_c(I)}{2},\frac{L+r_c(I)}{2})\\,(\frac{L+r_c(I)}{2}+1,\frac{L+r_c(I)}{2}+1))\, f$ as the effective capacity $c_{eff}(L/2, L/2)$.

With the effective capacity, we can obtain an analytical approximation for the average delay with background, yet it requires an extra consideration. In contrast to 1D lattices, it is not enough to replace $c$ by $c_{eff}$. Before entering a vehicle individuals walking through a link accumulate a delay of $\frac{q^{\star}}{c_{eff}}$, which corresponds to the time that the previous nodes take to recover. In 2D lattices, it only holds for individuals following the main axes along their route since they need to wait until the previous nodes have emptied. If, instead, they change direction after moving along the main axis, their delay will be different. Effectively they will accumulate $\frac{q^{\star}}{c}$ in the first walking direction, in which they do not take any vehicle, and $\frac{q^{\star}}{c_eff}$ in the final direction, in which they take a vehicle. In other words, if an individual takes a vehicle upwards in a node at a distance $3$ to the right and $1$ up of the origin, his delay induced by waiting the vehicle will be given by $f(\frac{q^{\star}}{c_eff}+3\frac{q^{\star}}{c})$. Taking this fact into account the total delay of the individuals that wait for a vehicle when there is background yields
\begin{align}\label{bet2db}
&\Delta \tau_{tot,v}=(\frac{q^2}{2c_{eff}}+\frac{q^2}{2c})f\frac{8}{3}(r_c-2)(r_c-1)r_c \\  
& +6(-1+r_c)r_cfq^2(\frac{1}{3c_eff}+\frac{2}{3c}) \nonumber \\ & + f \, p \, \left( \frac{c\, (p-1)}{2} + (q^{\star}- p\, c) \right) \, 4\, r_c^2 . \notag
\end{align}

The first term accounts for the delay of the individuals out of the main axes, whose delay is the mean between $\frac{q^{\star}}{c_eff}$ and $\frac{q^{\star}}{c}$. The second term obeys to the delay of the individuals in the main axes, whose delay per link is given by $q^{\star}(\frac{1}{3c_eff}+\frac{2}{3c})$ which is a consequence of one parallel direction of emptying, which gives $\frac{q^{\star}}{c_eff}$ and two perpendicular not influenced by the effective capacity, and give $\frac{q^{\star}}{c}$. Both the simulations and the analytical solution are shown in Fig. \ref{Figure10}, displaying a good agreement.

\begin{figure}[ht]
\begin{center}
\includegraphics[width=8cm]{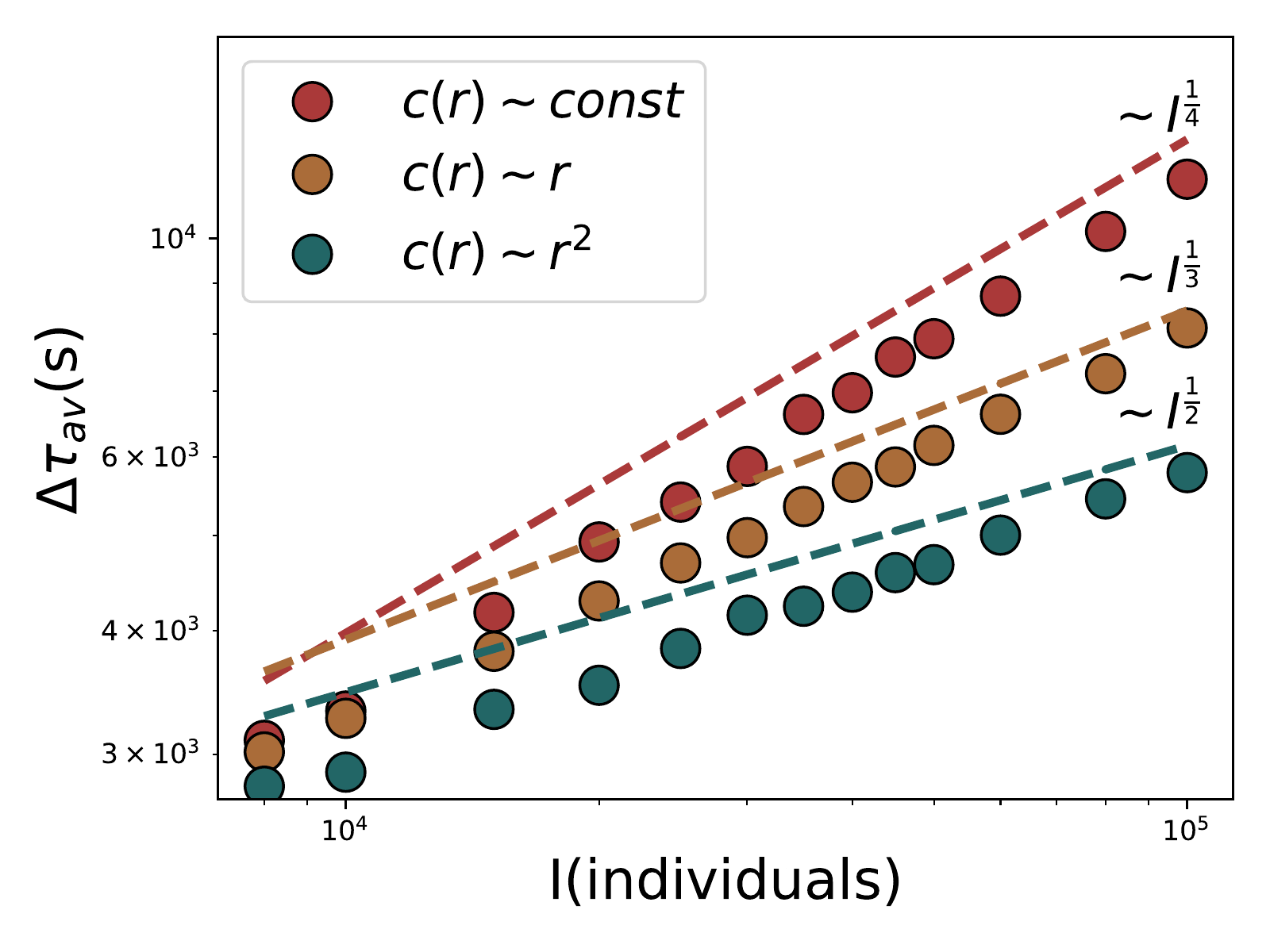}
\caption[Scaling of the average delay with $I$ in a 2D lattice with heterogeneous capacities]{Scaling of $\Delta \tau_{av}$ with $I$ in a 2D lattice with a capacity which depends on the distance from the center of the lattice where the event is introduced.
\label{Figure11}}
\end{center}
\end{figure}

\begin{figure}[ht]
\begin{center}
\includegraphics[width=8cm]{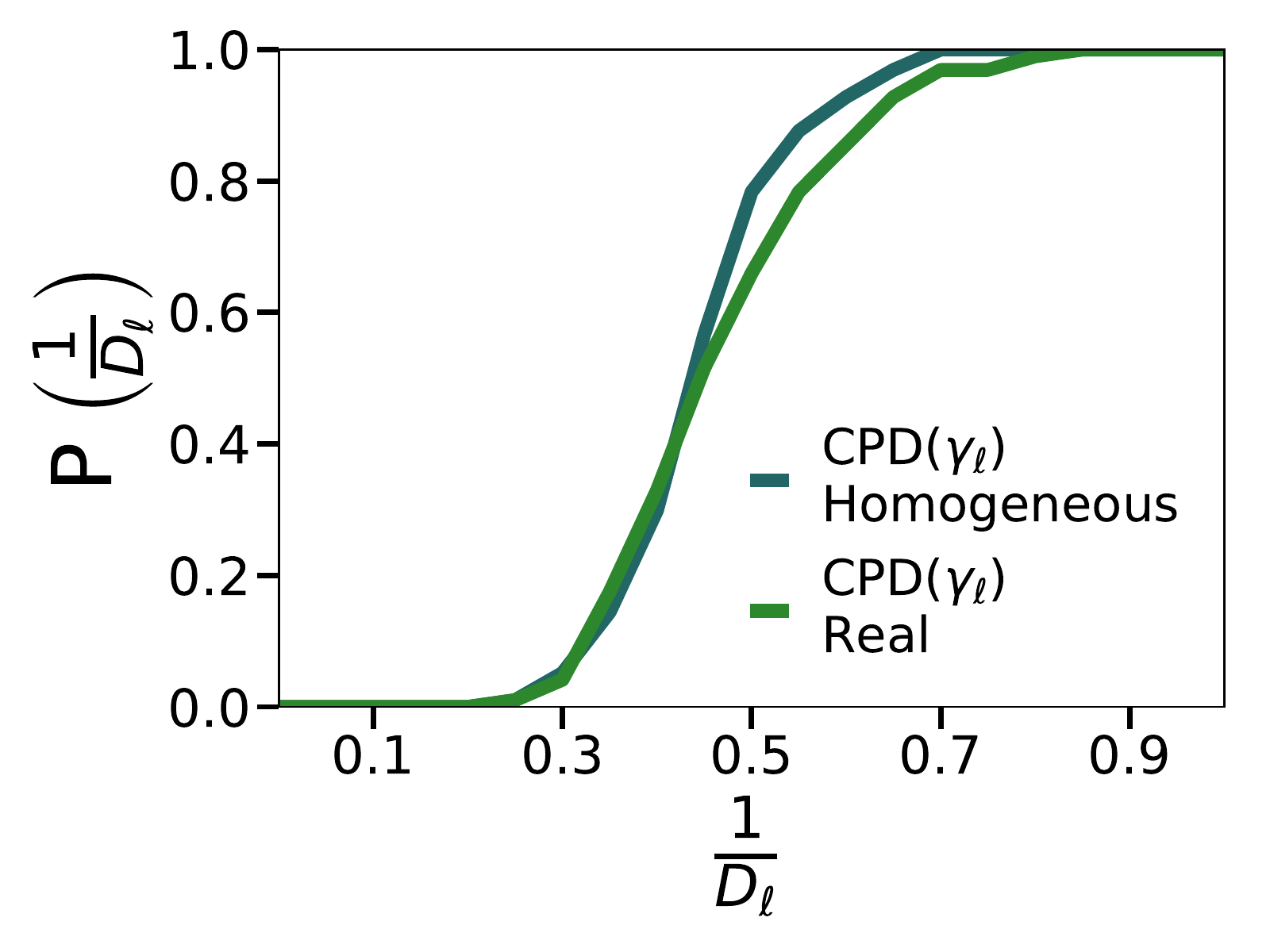}
\caption[Cumulative probability distribution of scaling exponents in Paris with real and homogeneous characteristics.]{Cumulative probability distribution of scaling exponents in Paris with the real and homogeneous speeds, capacities and periods.
\label{Figure12}}
\end{center}
\end{figure}
\section{Scaling with inhomogeneous capacity distribution}

Here we address the effect of inhomogeneous capacities in both regular lattices and cities. In the case of lattices, we show that if the capacities of the links are modified, while keeping the 2D topology, a scaling below $1/2$ appears. We will consider three situations: a constant capacity across the network, a linear and a quadratic increase with the distance from the event. Similarly to the previous sections, we calculate the scaling of the average delay $\Delta \tau_{av}$ with the event individuals in each of the cases (Fig. \ref{Figure11}). In the case of constant capacity, we observe the already stated scaling. While it is modified in the other two cases: If the capacity increases linearly with the distance to the event, the scaling approaches $1/3$. Otherwise, if it increases quadratically with distance, the scaling of the delay approaches $1/4$.

In the main manuscript, we discuss that the broad exponent distribution in cities is a consequence of the different speeds of each transportation modality, the inhomogeneous capacities and frequencies. We performed simulations in Paris keeping all the lines as they are in the map, but all mode speeds to $30 km/h$, a period of $10$ minutes and a capacity of $200$ persons. In Fig. \ref{Figure12}, we compare the cumulative distribution of exponents for both the real network and the homogenized one. As can be seen, the distribution of exponents is much more peaked around $0.5$ when the characteristics of all lines are homogeneous.

\section{Results in other cities}

\begin{figure}[b]
\begin{center}
\includegraphics[width=4.2cm]{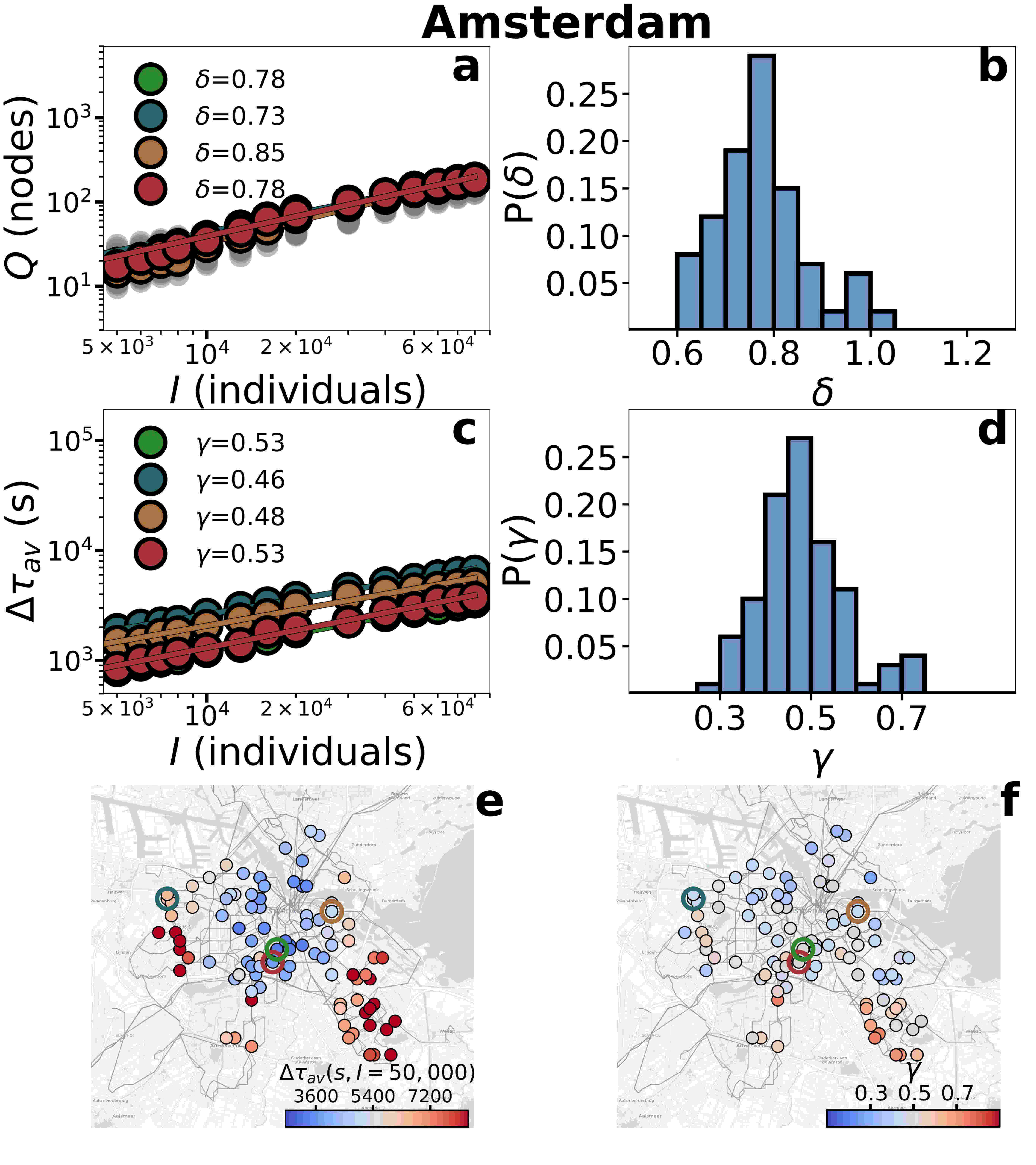}
\includegraphics[width=4.2cm]{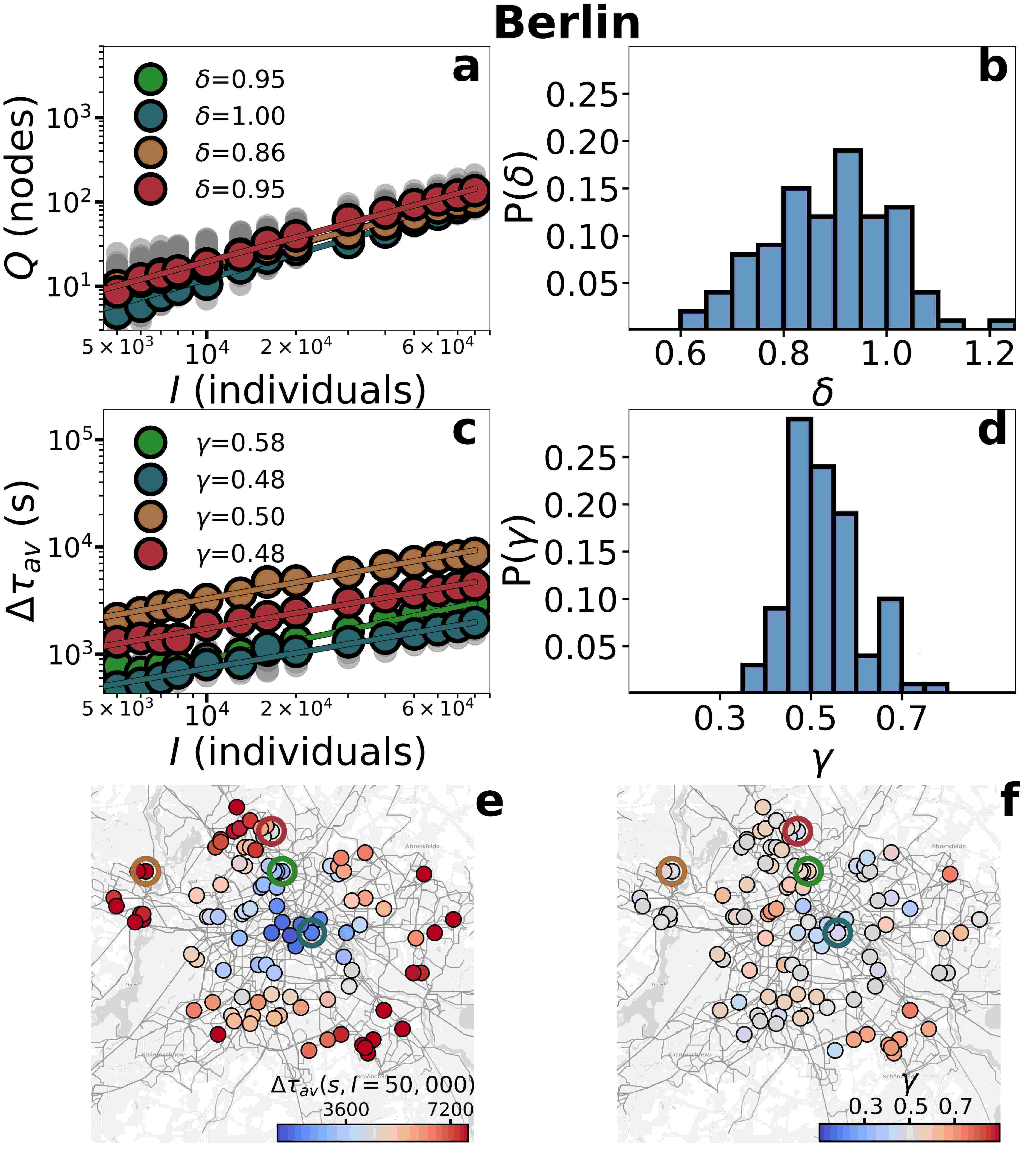}
\caption[Scaling for the event individuals in Amsterdam and Berlin]{Scaling for the event individuals in Amsterdam and Berlin. \textbf{a} Scaling of congested nodes with $I$. \textbf{b} Distribution of the exponents of the scaling of congested nodes. \textbf{c} Scaling of the average delay with $I$. \textbf{d} Distribution of the exponents of the scaling of the average delay. \textbf{e} Map of the scaling exponents of the average delay. \textbf{f} Map of the average delay for an event of $50,000$ individuals.The empty circles in the maps mark the locations of the scaling.
\label{FigureS1}}
\end{center}
\end{figure}

\begin{figure}[H]
\includegraphics[width=4.2cm]{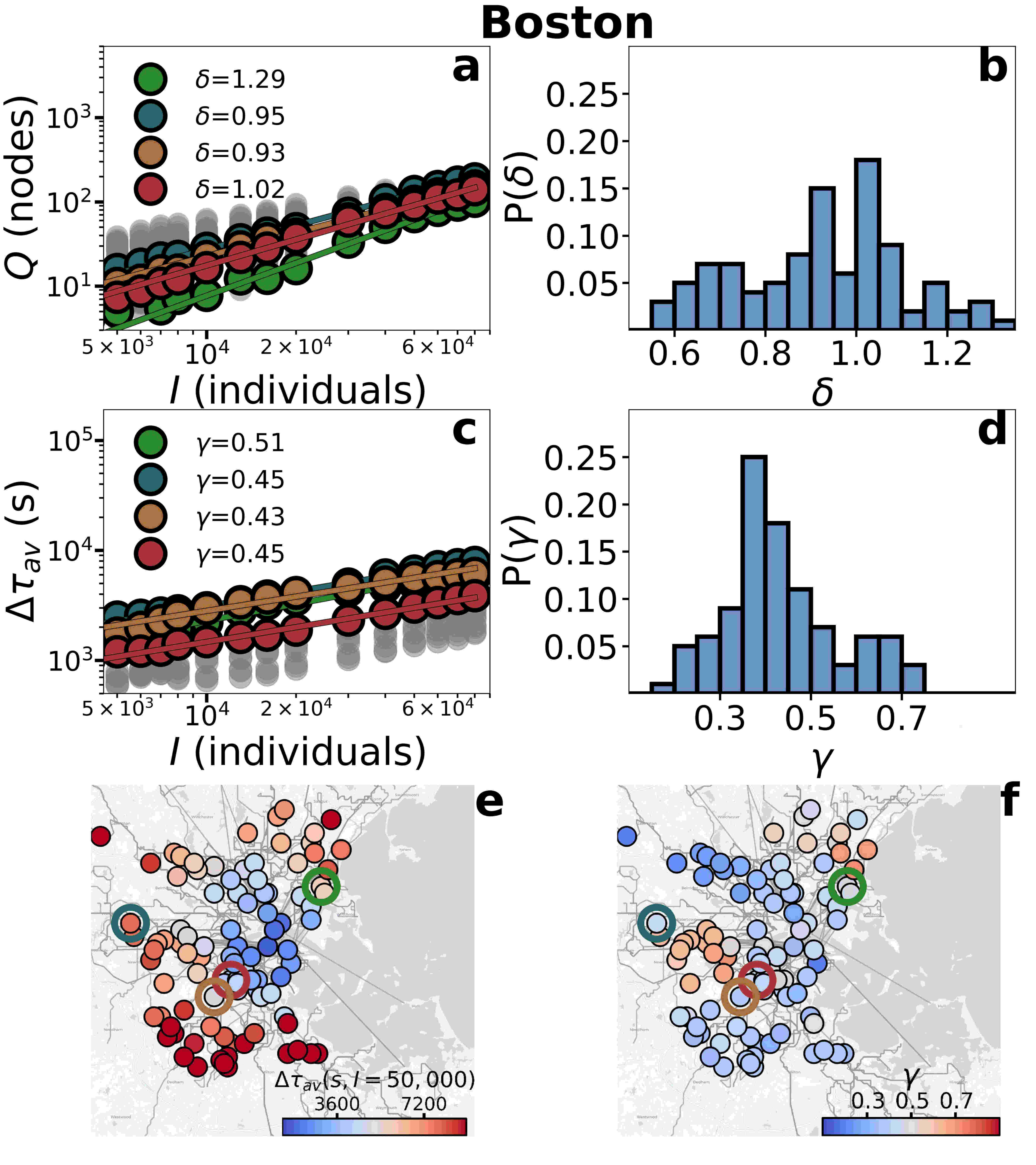}
\includegraphics[width=4.2cm]{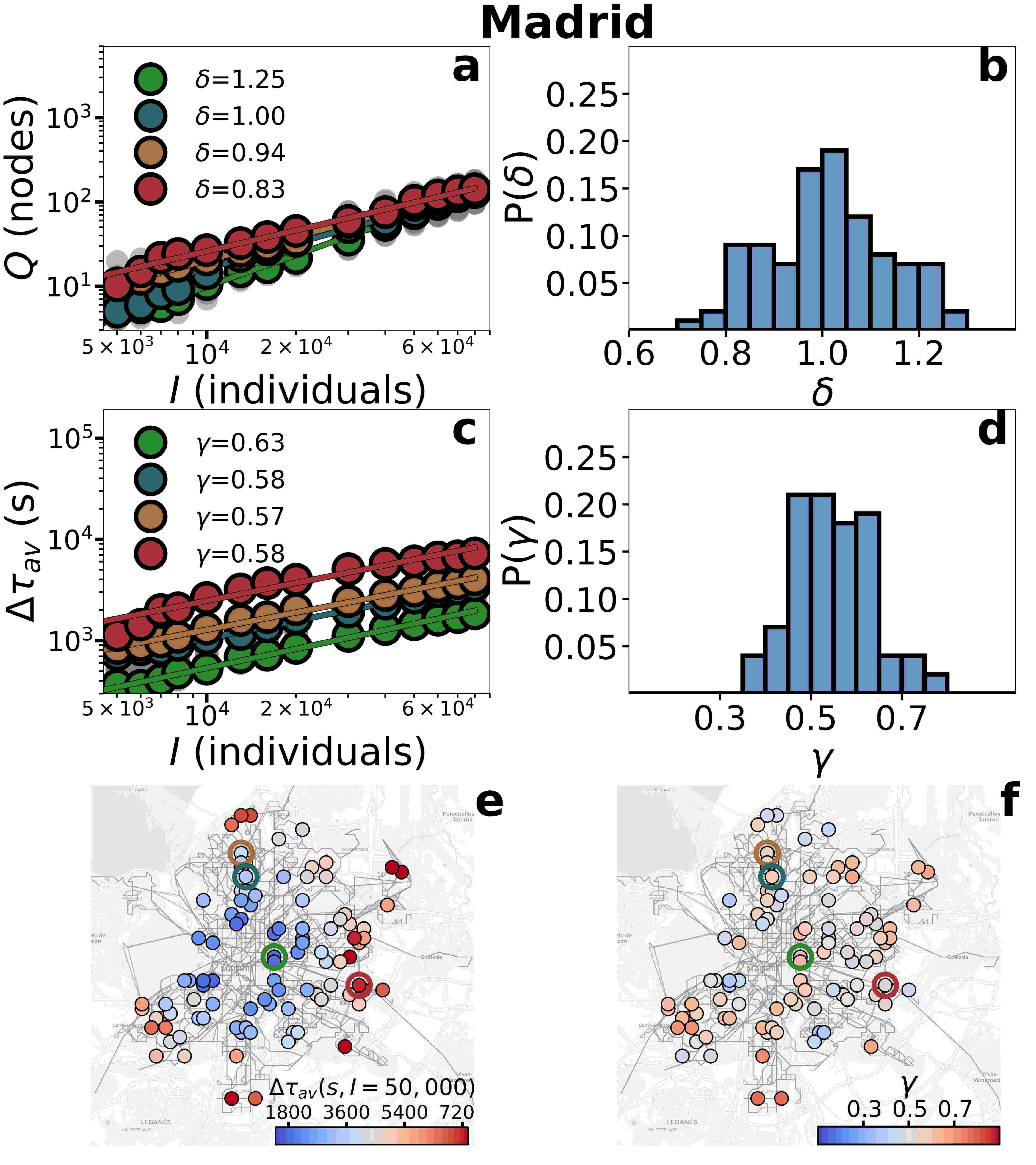}
\caption[Scaling for the event individuals in Boston and Madrid]{Scaling for the event individuals in Boston and Madrid.  \textbf{a} Scaling of congested nodes with $I$. \textbf{b} Distribution of the exponents of the scaling of congested nodes. \textbf{c} Scaling of the average delay with $I$. \textbf{d} Distribution of the exponents of the scaling of the average delay. \textbf{e} Map of the scaling exponents of the average delay. \textbf{f} Map of the average delay for an event of $50,000$ individuals.The empty circles in the maps mark the locations of the scaling.
\label{FigureS2}}
\includegraphics[width=4.2cm]{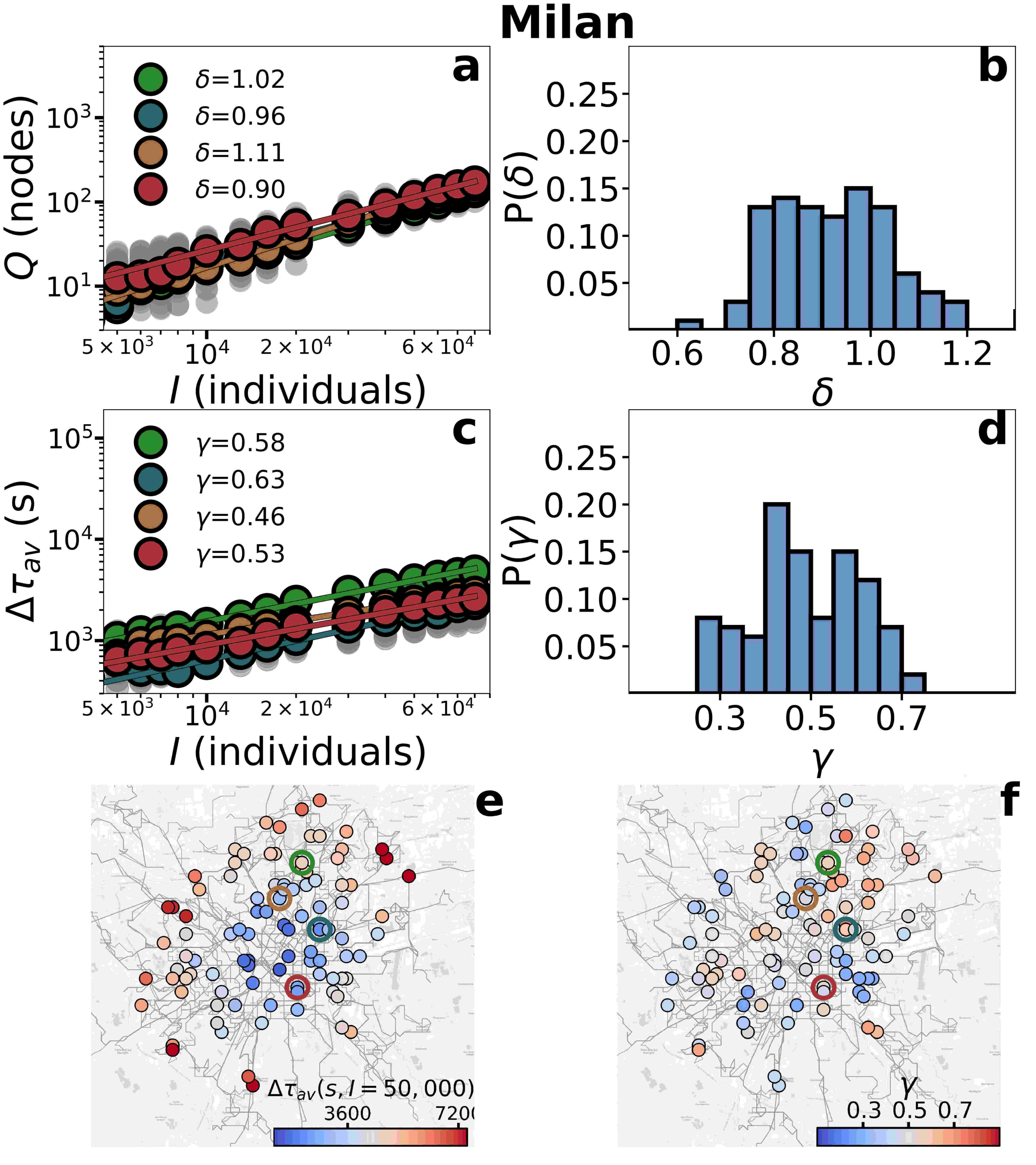}
\includegraphics[width=4.2cm]{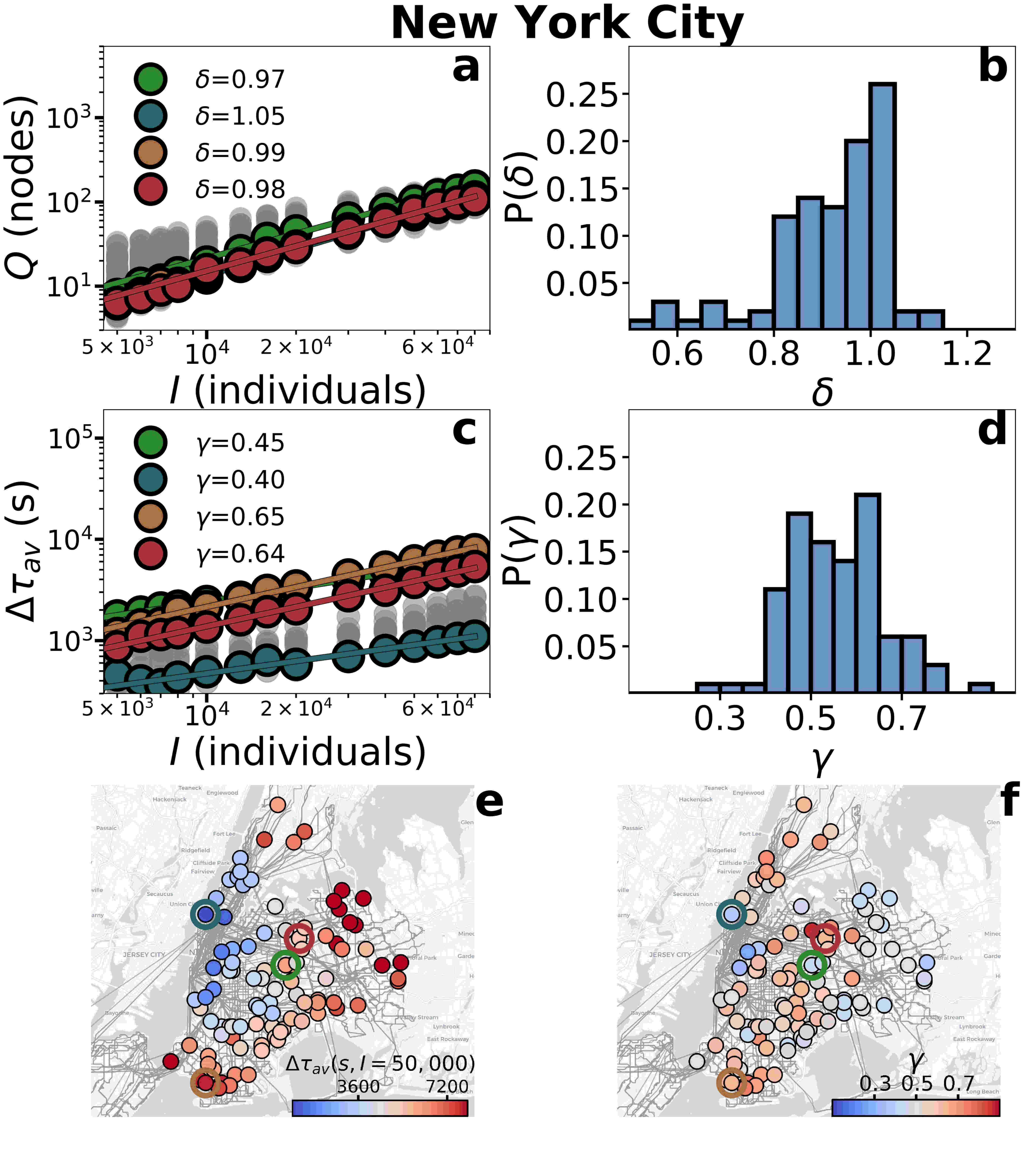}
\caption[Scaling for the event individuals in Milan and New York City]{Scaling for the event individuals in Milan and New York City.  \textbf{a} Scaling of congested nodes with $I$. \textbf{b} Distribution of the exponents of the scaling of congested nodes. \textbf{c} Scaling of the average delay with $I$. \textbf{d} Distribution of the exponents of the scaling of the average delay. \textbf{e} Map of the scaling exponents of the average delay. \textbf{f} Map of the average delay for an event of $50,000$ individuals.The empty circles in the maps mark the locations of the scaling.
\label{FigureS3}}
\end{figure}

\begin{figure}[H]
\begin{center}
\includegraphics[width=4.2cm]{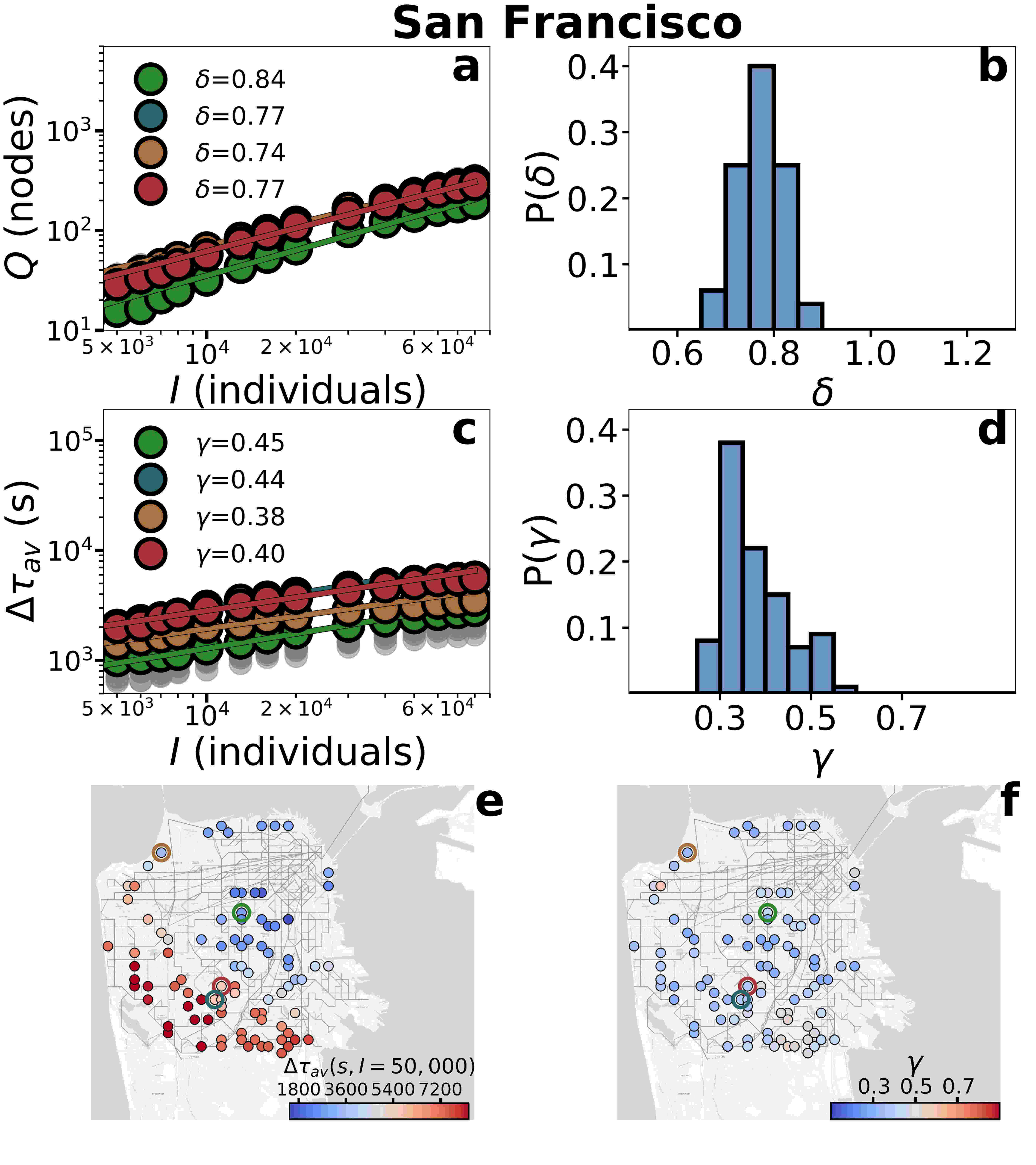}
\caption[Scaling for the event individuals in San Francisco]{Scaling for the event individuals in San Francisco.  \textbf{a} Scaling of congested nodes with $I$. \textbf{b} Distribution of the exponents of the scaling of congested nodes. \textbf{c} Scaling of the average delay with $I$. \textbf{d} Distribution of the exponents of the scaling of the average delay. \textbf{e} Map of the scaling exponents of the average delay. \textbf{f} Map of the average delay for an event of $50,000$ individuals. The deviation in the exponent distribution from $0.5$ is due to the finite size of the area considered. Even the scaling of the variables of the panels (a) and (b) are not of so high quality in this case, with the curves near a plateau for large $I$. The empty circles in the maps mark the locations of the scaling.
\label{FigureS4}}
\includegraphics[width=8.4cm]{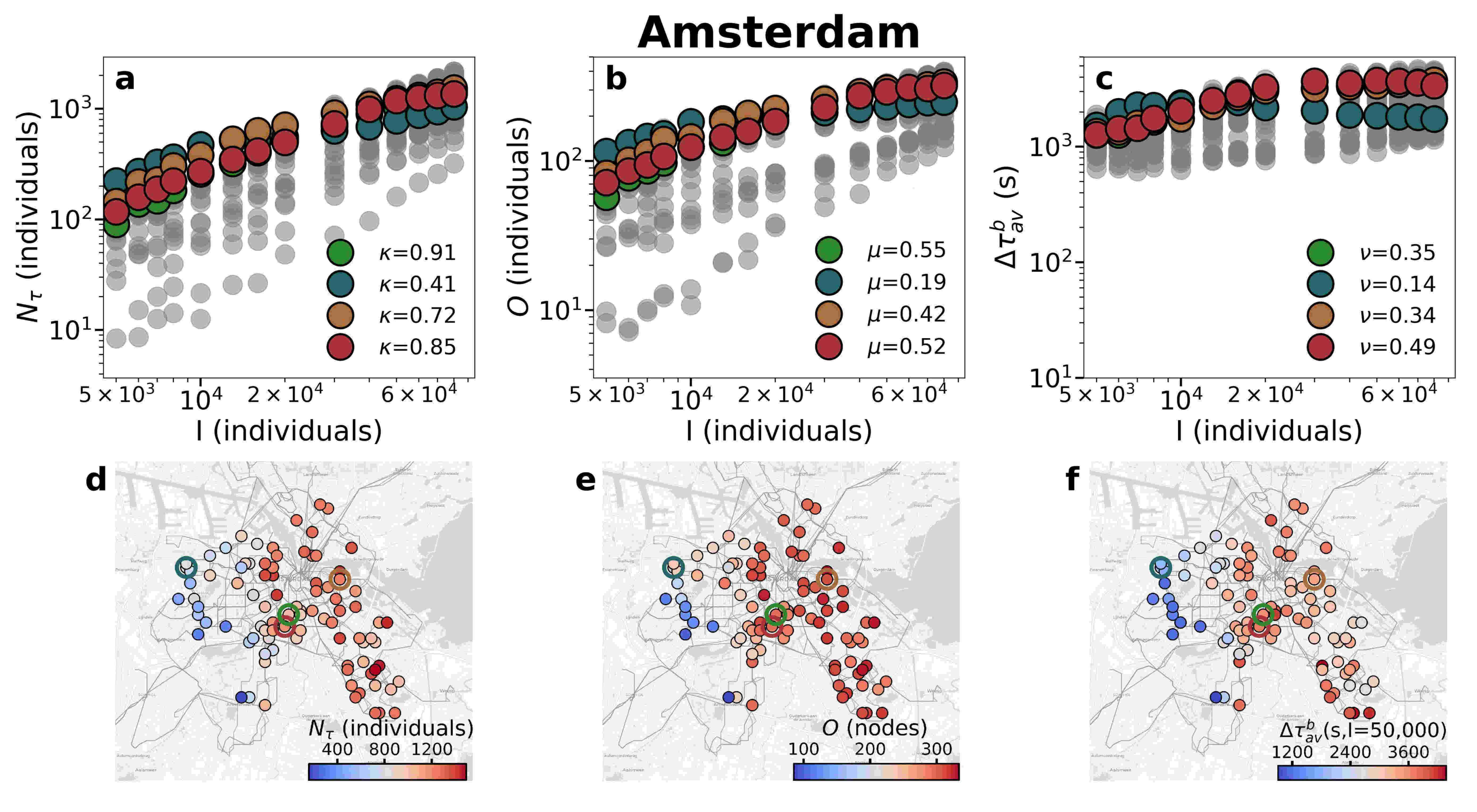}
\caption[Scaling for the background individuals in Amsterdam]{Scaling for the background individuals in Amsterdam. Scaling of the \textbf{a} delayed individuals, \textbf{b} origins affected and \textbf{c} average delay with the number event individuals. Map of the \textbf{d} delayed individuals, \textbf{e} origins affected and \textbf{f} average delay for an event of $50,000$ individuals. The empty circles in the maps mark the locations of the scaling shown in \textbf{a}, \textbf{b} and \textbf{c}. The empty circles in the maps mark the locations of the scaling.
\label{FigureS7}}
\end{center}
\end{figure}

\begin{figure}[H]
\begin{center}
\includegraphics[width=8.4cm]{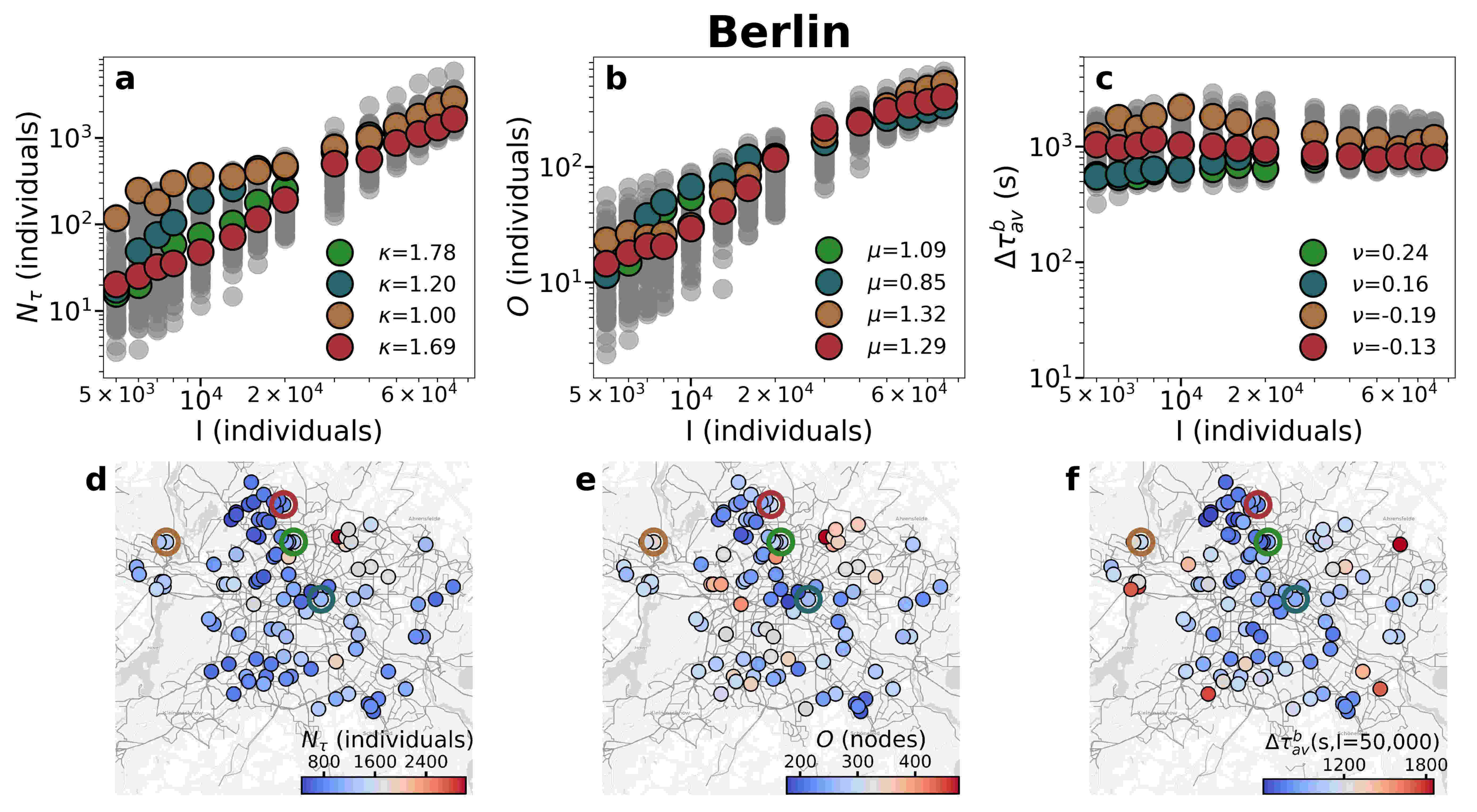}
\caption[Scaling for the background individuals in Berlin]{Scaling for the background individuals in Berlin. Scaling of the \textbf{a} delayed individuals, \textbf{b} origins affected and \textbf{c} average delay with the number event individuals. Map of the \textbf{d} delayed individuals, \textbf{e} origins affected and \textbf{f} average delay for an event of $50,000$ individuals. The empty circles in the maps mark the locations of the scaling shown in \textbf{a}, \textbf{b} and \textbf{c}. The empty circles in the maps mark the locations of the scaling.
\label{FigureS8}}
\includegraphics[width=8.4cm]{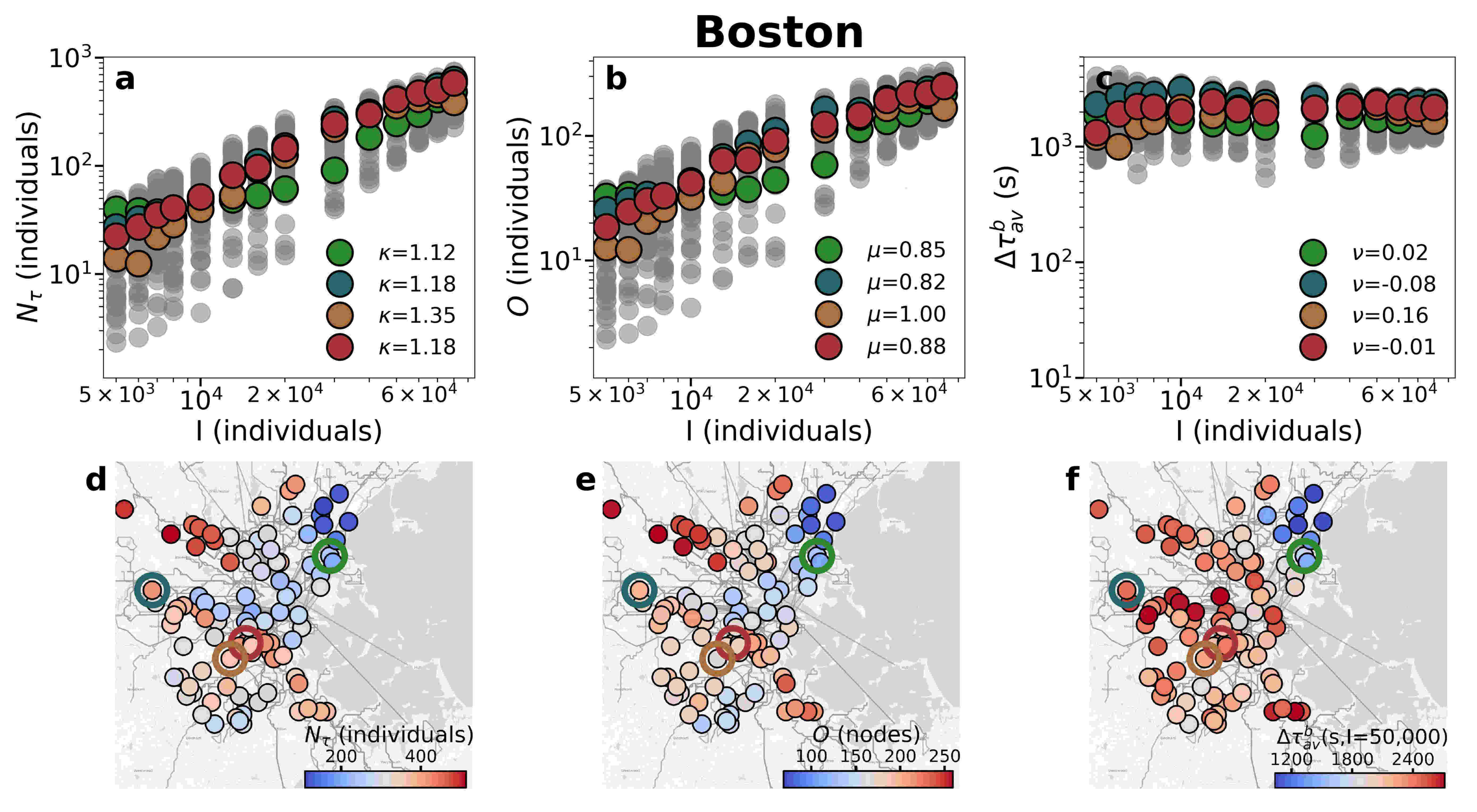}
\caption[Scaling for the background individuals in Boston]{Scaling for the background individuals in Boston. Scaling of the \textbf{a} delayed individuals, \textbf{b} origins affected and \textbf{c} average delay with the number event individuals. Map of the \textbf{d} delayed individuals, \textbf{e} origins affected and \textbf{f} average delay for an event of $50,000$ individuals. The empty circles in the maps mark the locations of the scaling shown in \textbf{a}, \textbf{b} and \textbf{c}. The empty circles in the maps mark the locations of the scaling.\label{FigureS9}}
\includegraphics[width=8.4cm]{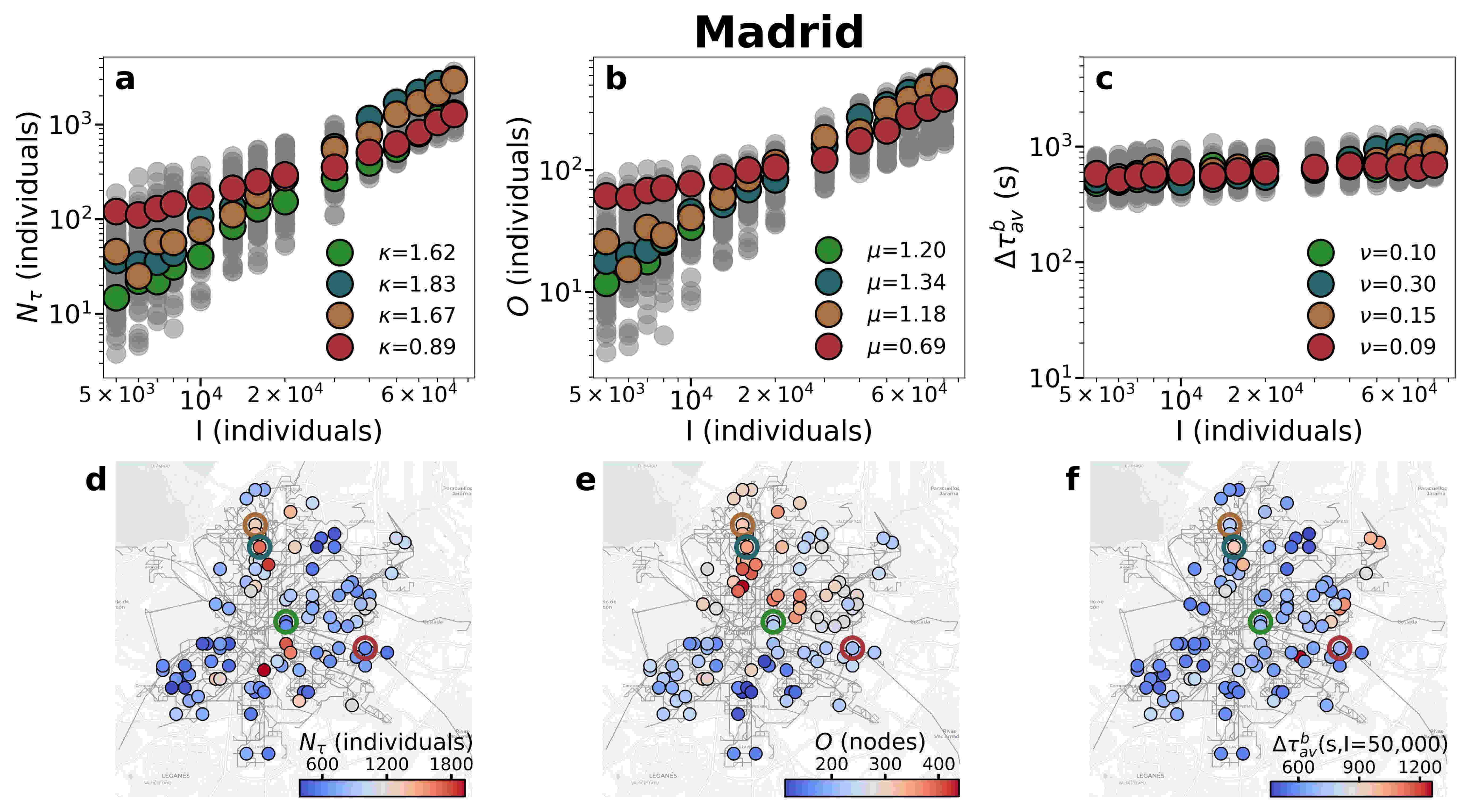}
\caption[Scaling for the background individuals in Madrid]{Scaling for the background individuals in Madrid. Scaling of the \textbf{a} delayed individuals, \textbf{b} origins affected and \textbf{c} average delay with the number event individuals. Map of the \textbf{d} delayed individuals, \textbf{e} origins affected and \textbf{f} average delay for an event of $50,000$ individuals. The empty circles in the maps mark the locations of the scaling shown in \textbf{a}, \textbf{b} and \textbf{c}. The empty circles in the maps mark the locations of the scaling.
\label{FigureS10}}
\end{center}
\end{figure}

\begin{figure}[H]
\begin{center}
\includegraphics[width=8.4cm]{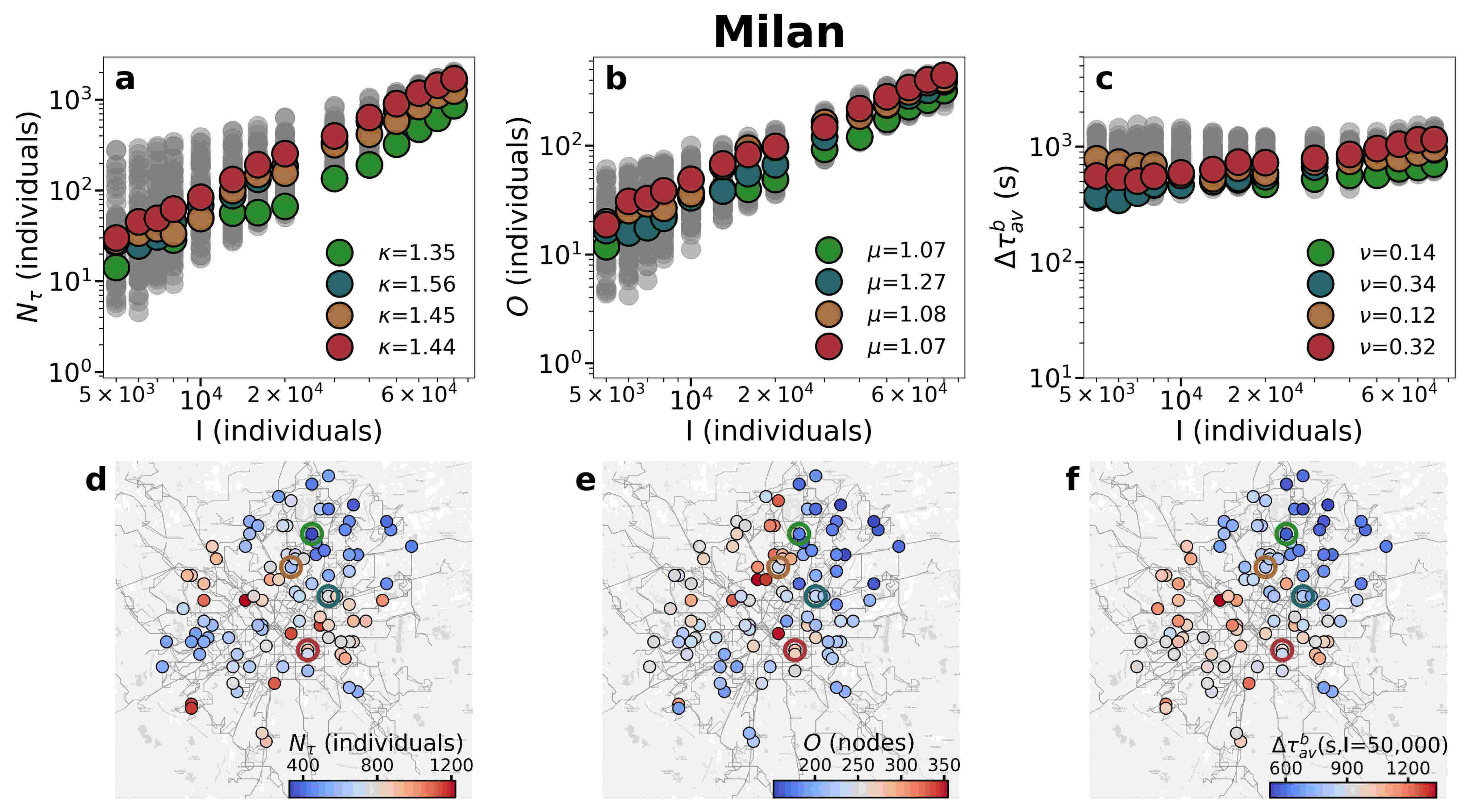}
\caption[Scaling for the background individuals in Milan]{Scaling for the background individuals in Milan. Scaling of the \textbf{a} delayed individuals, \textbf{b} origins affected and \textbf{c} average delay with the number event individuals. Map of the \textbf{d} delayed individuals, \textbf{e} origins affected and \textbf{f} average delay for an event of $50,000$ individuals. The empty circles in the maps mark the locations of the scaling shown in \textbf{a}, \textbf{b} and \textbf{c}. The empty circles in the maps mark the locations of the scaling.
\label{FigureS11}}
\includegraphics[width=8.4cm]{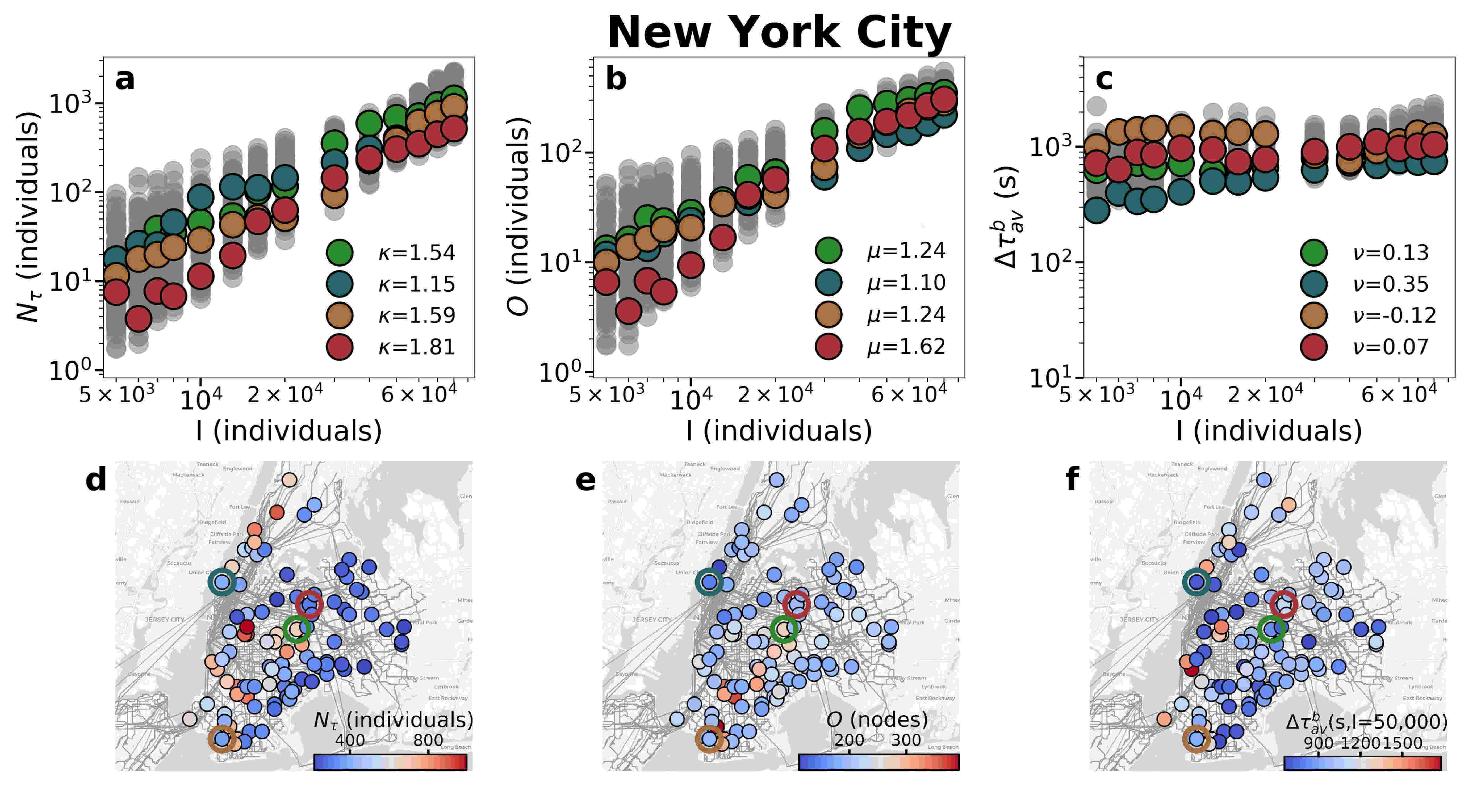}
\caption[Scaling for the background individuals in New York City]{Scaling for the background individuals in New York City.Scaling of the \textbf{a} delayed individuals, \textbf{b} origins affected and \textbf{c} average delay with the number event individuals. Map of the \textbf{d} delayed individuals, \textbf{e} origins affected and \textbf{f} average delay for an event of $50,000$ individuals. The empty circles in the maps mark the locations of the scaling shown in \textbf{a}, \textbf{b} and \textbf{c}. The empty circles in the maps mark the locations of the scaling.
\label{FigureS12}}
\includegraphics[width=8.4cm]{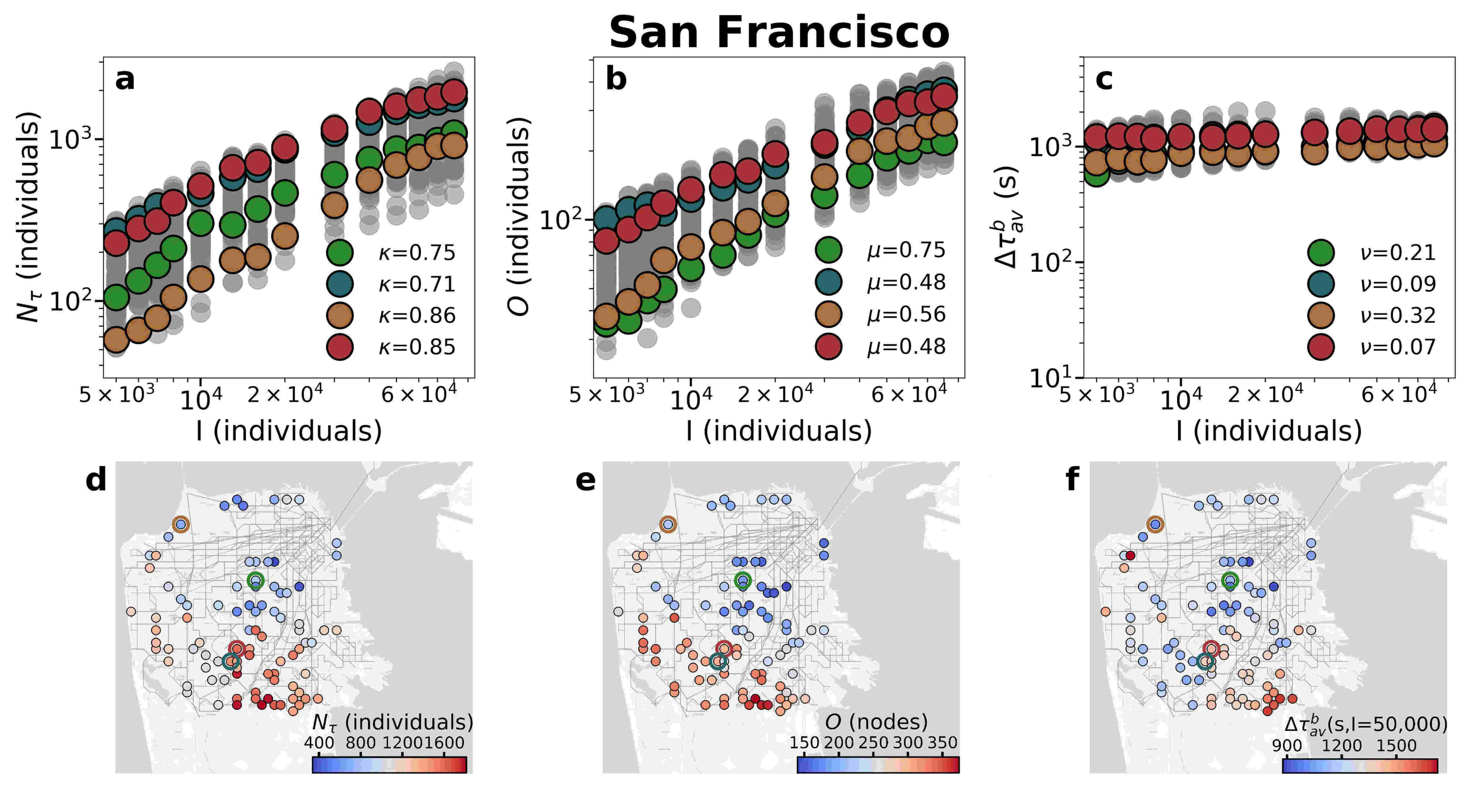}
\caption[Scaling for the background individuals in San Francisco]{Scaling for the background individuals in San Francisco. Scaling of the \textbf{a} delayed individuals, \textbf{b} origins affected and \textbf{c} average delay with the number event individuals. Map of the \textbf{d} delayed individuals, \textbf{e} origins affected and \textbf{f} average delay for an event of $50,000$ individuals. The empty circles in the maps mark the locations of the scaling shown in \textbf{a}, \textbf{b} and \textbf{c}. The empty circles in the maps mark the locations of the scaling.
\label{FigureS13}}
\end{center}
\end{figure}

\begin{figure*}
\begin{center}
\includegraphics[width=14cm]{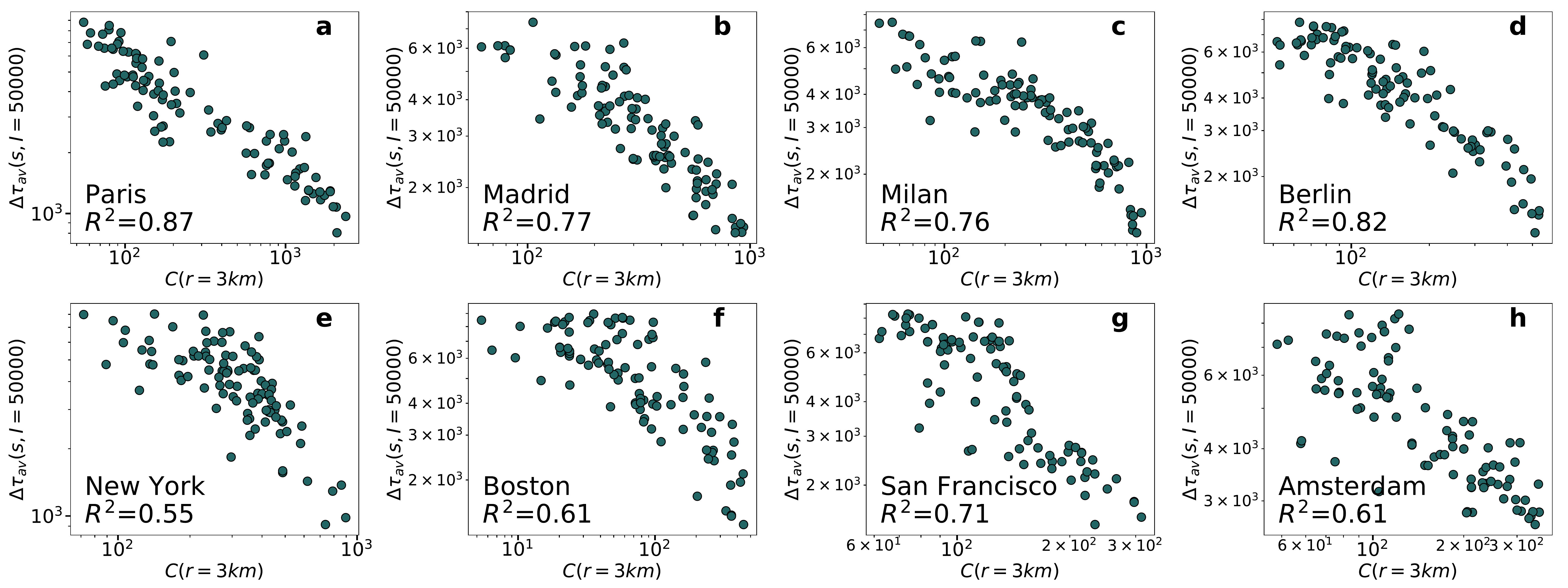}
\caption[Comparing the local capacity $C(r)$ and the average delay $\Delta\tau_{av}$]{Comparing the total capacity $C(r)$ within a radius of $3$ km and the average delay $\Delta\tau_{av}$. $\Delta\tau_{av}$ for an event with $50,000$ individuals as a function of the total capacity within a radius of $3$ km for the eight cities of study: \textbf{a} Paris, \textbf{b} Madrid, \textbf{c} Milan, \textbf{d} Berlin, \textbf{e} New York City, \textbf{f} Boston, \textbf{g} San Francisco, \textbf{h} Amsterdam.
\label{FigureS5}}
\end{center}
\end{figure*}

In this section, we provide the scaling results obtained in the other seven cities. Figures \ref{FigureS1}-\ref{FigureS4} depict the results of the scaling for the event individuals in Amsterdam, Berlin, Boston, Madrid, Milan, New York City and San Francisco. Figure \ref{FigureS5} provides the correlation between $\Delta \tau_{av}(I)$ for $I=50,000$ and the total capacity in a radius of $3$km for a set of 100 locations in the cities of study. Finally, we examine the effect of the event location on the background, Figures \ref{FigureS7}-\ref{FigureS13} display the results of the scaling for the individuals in the background.

\end{document}